# Eleven Years of QCD at LEP

S. Bethke

Max-Planck-Institute of Physics, 80805 Munich, Germany
e-mail: bethke@mppmu.mpg.de

*In memory of Siegfried Weseler*

**Abstract.** Studies of hadronic final states of $e^+e^-$ annihilations, observed at the Large Electron Positron Collider LEP at CERN, are reviewed. The topics included cover measurements of $\alpha_s$, hadronic event shapes and hadronisation studies, tests of asymptotic freedom and of the non-Abelian gauge structure of QCD, differences between quark and gluon jets, tests of power corrections and selected results of two-photon scattering processes. The improvements obtained at LEP are demonstrated by comparing to results from the pre-LEP era. This article consists of a reproduction of slides presented at the LEPFest in October 2000, supplemented by a short descriptive text and a list of relevant references.

PACS: not given

## 1 Introduction

Between the initial start-up of the electron-positron collider LEP [1] in August 1989 and its close-down in November 2000, the four particle physics experiments ALEPH [2], DELPHI [3], L3 [4] and OPAL [5] have contributed more than 200 publications on hadronic physics and tests of Quantum Chromodynamics (QCD), the theory of the Strong Interaction between quarks and gluons. This article gives an overview of the status of QCD analyses and tests which were available at the time of the close-down of LEP. It is based on a presentation given at the LEPFest in October 2000, celebrating more than 11 years of successful data taking.

A general introduction to the physics and tests of QCD at particle colliders can be found e.g. in [6]. Earlier reviews of QCD results from LEP are available in [7, 8, 9]. Summaries of the performance of LEP and of the 4 experiments, and further reviews of physics results obtained at LEP were reported at the LEPFest. They can be accessed at the relevant web pages at CERN [10].

The topics covered in this presentation are given on **Slide 1**, along with an excuse about those topics which - due to time and space limitations - could not be included in this review. **Slide 1** also contains an acknowledgement of individuals, groups and organisations whose contributions to the physics at LEP and to this review were invaluable.

## 2 Hadronic Events at LEP

The general anatomy of a hadronic event in electron positron annihilation is illustrated on **Slide 2**. The development of a quark-gluon cascade from the initial quark-antiquark system can be calculated in fixed order QCD perturbation theory, nowadays available in full next-to-leading order (NLO, equivalent to $\mathcal{O}(\alpha_s^2)$) [11, 12, 13], or in the (next-to-)leading logarithmic approximation ((N)LLA) [14]. The nonperturbative process of hadronisation into visible particles is described by either string- [15] or cluster- [16] fragmentation models or, alternatively, by applying analytical power corrections [17], see also **Slides 19 and 20**.

Illustrative examples of hadronic final states as measured with the OPAL detector at LEP-I, at a centre-of-mass energy of 91.2 GeV (i.e. on the $Z^0$ pole) and close to the highest energies reached at LEP-II, at Ecm=205.4 GeV, are given in **Slide 2** and in **Slide 3**, respectively. At these energies, multi-jet structures are clearly visible from the reconstructed tracks of charged particles as well as from the energy flows measured in the electromagnetic and hadronic calorimeters.

Significant improvements in tests of QCD and in hadronic physics, in general, were obtained at LEP, as will be demonstrated in the following by comparing the actual status of some topics to the knowledge available in 1989, before the start-up of LEP. These improvements were possible due to the items discussed on **Slide 4**.

## 3  $\alpha_s$ in $e^+e^-$ Annihilations

The status of measurements of $\alpha_s$ in $e^+e^-$ annihilation, in the year 1989, before the startup of LEP, and in 2000, is summarised on **Slide 5**. In 1989, several new analyses became available from the experiments at the PETRA and PEP $e^+e^-$ colliders. There, $\alpha_s$ was determined from event shape observables, from energy correlelations, from the normalised total hadronic cross section R and from 3-jet event rates [18]. The analyses were based on different analytic QCD calculations in next-to-leading order [11, 19] and relied on different models to correct for hadronisation effects, namely the independent jet fragmentation and the string fragmentation model [20]. The results revealed systematic biases from different hadronisation models, but also from the use of different observables and different theoretical calculations in NLO QCD. The latter two effects are nowadays known as theoretical uncertainties of the calculations in truncated order of perturbation theory, which affect different theoretical approaches and different observables in a different way. The observed systematic biases limited the precision of a combined result to a value [18] of

$$\begin{aligned}\alpha_s(35\,\text{GeV}) &= 0.14 \pm 0.02\,, \quad \text{corresponding to} \\ \alpha_s(M_{Z^0}) &= 0.119 \pm 0.016\,.\end{aligned}$$

Actual values of $\alpha_s$ from $e^+e^-$ annihilations in the year 2000, at the end of the LEP running period, are summarised [21] on the right-hand side of **Slide 5**. All results, obtained in the energy range from the $\tau$ lepton rest mass, $M_\tau = 1.78$ GeV, up to the highest energies reached at LEP, are converted to $\alpha_s(M_{Z^0})$. Results with filled symbols are based on complete next-next-to-leading order perturbation theory (NNLO or $\mathcal{O}(\alpha_s^3)$) [22], those with open symbols are based on resummed [14] next-to-leading order calculations [11, 12, 13] of hadronic event shapes and jet rates. Symbols in red are results obtained at LEP, those in green are from other $e^+e^-$ collider experiments. For a comprehensive reference to the data summarised in this figure, see [21]. The results for

c.m. energies of 200 and 206 GeV are not yet published and are preliminary. Averaging the results shown in this figure, separately for those in resummed NLO and in full NNLO QCD, gives

$$\alpha_s(M_{Z^0}) = 0.121 \pm 0.005 \quad \text{(resummed NLO), and}$$
$$\alpha_s(M_{Z^0}) = 0.120 \pm 0.003 \quad \text{(NNLO)}.$$

Here, a method to combine results which are subject to an unknown degree of correlated errors [23] was used; more detailed analyses aiming at a consistent treatment and combination of all LEP results on $\alpha_s$ are currently prepared by the LEP QCD Study Group.

## 4 World Summary of $\alpha_s$

A world summary of $\alpha_s$ is presented on **Slide 6**. On the left-hand side, the situation as summarised in 1989 [24], prior to the start-up of LEP, is reproduced. At that time, these data could not prove that $\alpha_s$ is running. A fit of QCD to the data predicted, however, that $\alpha_s(M_{Z^0}) = 0.110 \pm 0.007$ (when symmetrising the error).

The status of measurements of $\alpha_s$ has significantly changed until the year 2000, as is demonstrated by the figure on the right-hand-side [21]: precise results at low as well as at high energies, from the mass scale of the $\tau$-lepton up to the highest LEP-II energies, convincingly demonstrate the running of $\alpha_s$, in perfect agreement with QCD. They result in a world average value of

$$\alpha_s(M_{Z^0}) = 0.1184 \pm 0.0031 \quad \text{(NNLO)}, \tag{1}$$

whereby, for the first time, a sufficient number of results based on complete NNLO QCD was available to provide a precise, combined value of $\alpha_s$ [21].

## 5 Asymptotic Freedom from Jet Rates

Asymptotic freedom, i.e. the running of $\alpha_s$, was demonstrated in studies of 3-jet production rates, $R_3$ [25], long before measurements of $\alpha_s$ provided a significant case. The JADE-E0 jet finder [26] was found to be subject to small and almost energy independent hadronisation corrections, see the left figure on **Slide 7**. Measurements of $R_3$, which in leading order QCD is exactly proportional to $\alpha_s$, provided early indications for the running of $\alpha_s$, especially with the advent of the LEP data. They fell exactly on the expectations from lower energy data assuming the validity of QCD (right-hand side figure on **Slide 7**) [27]. The same data, combined at similar energies and with the LEP-II results added, are displayed on **Slide 8**, now as a function of $\ln(1/E_{cm})$. They are compatible with the QCD prediction of asymptotic freedom which, in leading order, is a straight line running through the origin of this graph, i.e. $\alpha_s \to 0$ as $E_{cm} \to \infty$.

## 6 Non-Abelian Gauge Structure from 4-Jet Events

As an alternative to test the non-Abelian gauge structure of QCD through the running of $\alpha_s$, angular correlations within 4-jet events, which are sensitive to the existence of the gluon self-coupling, were studied. **Slide 9** demonstrates the increased analysing power of such studies at LEP. To the left, the distribution of the Bengtson-Zerwas angle [28] between the energy-ordered jet axes of reconstructed 4-jet events, as measured by the AMY collaboration at $e^+e^-$ collision energies around 56 GeV [29], is compared with the predictions of QCD and with an Abelian theory where the gluon self-coupling does not exist. To the right, the status of such a measurement after one year of data taking at LEP is displayed [30]. The current state-of-the art of such studies, which usually involve the analysis of several angular correlations, is summarised [31] on **Slide 10**. The data are in excellent agreement with the gauge structure constants of QCD ($C_A \equiv N_C = 3$, $C_F = 4/3$ and $T_R = 1/2$), and rule out the Abelian vector gluon model ($C_A = 0$, $C_F = 1$ and $T_R = 6$). Also the possible existence of light colour-charged spin-1/2 supersymmetric partners of the gluon, the gluinos, is strongly disfavoured.

## 7 Differences between q- and g-Jets

QCD predicts that quarks and gluons - due to their different colour charges - fragment differently, see the top of **Slide 11**: gluon initiated jets are expected to be broader than quark jets, the multiplicity of hadrons in gluon jets, $N_{had}$, should be larger than in quark jets, and particles in gluon jets are expected to be less energetic. The first of these conjectures was verified by JADE in 1982 [32], see the left part of **Slide 11**: the mean transverse momentum of particles with respect to the closest jet axis, $<p_t>$, was analysed as a function of the reconstructed jet energy $E_j$, in hadronic events which were classified as 3-jet events. The $<p_t>$ of the lowest energetic jet in such events, jet #3 which is most likely to originate from a gluon, was measured to be significantly larger than for jets #1 and 2 which most likely originate from the initial quark and antiquark. Hadronisation models assuming identical fragmentation of quarks and gluons predict a universal dependence of $<p_t>$ from the jet energy, see graph b) to the lower left.

At LEP, studies of differences between quark and gluon jets were further refined, e.g. by anti-tagging gluon jets through the help of high resolution silicon vertex detectors [33], see the right part of **Slide 11**: if in symmetric 3-jet events one of the two lower energetic jets exhibits a secondary vertex detached from the primary vertex, it is most likely due to the decay of an initial b-quark or antiquark. The other low energetic jet then is anti-tagged to be the gluon jet, with a probability, given by Monte Carlo model studies, of about 80%. In the graph to the lower right, the normalised hadron multiplicity distribution for such gluon-tagged jets is compared to the unbiased multiplicity distribution of the two lower energetic jets, which consists of a mixture of 50% quark- and 50% gluon jets. Gluon jets in hadronic 3-jet events at LEP appear to exhibit larger hadron multiplicities than quark jets, however by a much smaller amount than the factor of 9/4 which is predicted by leading order QCD, for infinite jet energies.

Further studies, analysing gluon-inclusive jets recoiling against two other jets which are double-tagged to be a b-quark-antiquark system [34], provided an increased significance for larger hadron multiplicities in gluon-jets, approaching and eventually reaching the asymptotic QCD prediction of 9/4 = 2.25, see the left part of **Slide 12**. The result of an alternative study [35], determining the charged particle multiplicity of hypothetical gluon-gluon jet events, from measurements of symmetric 3-jet events at LEP and from multiplicities of average hadronic (quark-antiquark-) events in $e^+e^-$ annihilation, is shown on the right-hand-side of this slide. The marked difference of hadron multiplicities in gluon jets, together with corresponding theoretical predictions in terms of QCD [36, 37], provided a precise fit of the ratio $C_A/C_F = 2.22 \pm 0.11$, in perfect agreement with the QCD expectation of 2.25.

## 8 String-Effect and Hadronisation Models

The string-effect is another manifestation of the different colour charge of quarks and gluons, and thus, of the non-abelian nature of QCD. It was experimentally verified in 1983 [38]. Analysing the particle flow in the regions between jets of identified and energy-ordered 3-jet events, a depletion of particles between the two most energetic jets #1 and 2, i.e. between the primary quark- and antiquark jets, was observed, in agreement with the string-picture inspired by QCD (see upper left graph of **Slide 13**). At LEP, higher jet energies, more refined jet algorithms and higher event statistics allowed to study the string effect with higher precision and significance, see the lower left and upper right graphs [39] of **Slide 13** (note the reversed sense of plotting the normalised particle flow angle $\theta$ w.r.t. the upper left graph). The ratio of particles $R$, measured in the central 40% of the angular regions between jets number 1 and 3, and jets number 1 and 2, discriminates between different hadronisation models, see the lower right graph on **Slide 13**.

The string-effect is a classical example of gluon coherence effects in hadronic final states [40]. Gluon coherence effects, in general, have extensively been studied at LEP [39, 41].

## 9 Jet Production Rates and Hadronic Event Shapes

Measurements of jet production rates and of hadronic event shape parameters developed into precision tools to determine $\alpha_s$, to probe details of perturbative QCD predictions and to study hadronisation properties. The left graph on **Slide 14** shows the relative production rates of 2-, 3-, 4- and multijet events measured at LEP-I [42], using the Durham (D-) scheme jet algorithm [43] which exhibits small hadronisation corrections and which can be calculated in resummed next-to-leading order perturbative QCD. The graph to the right demonstrates the good description of distributions of the Thrust ($T$) event shape parameter, measured in a wide range of centre of mass energies, by the respective resummed QCD calculations [44].

Precision measurements of hadronic event shape distributions, as shown for the $1 - T$ distribution on **Slide 15**, allowed to study and optimise different variants of perturbative QCD predictions [45]. As known from previous studies of data at lower c.m. energies [46], NLO perturbative QCD provides a good description of the shape of data distributions only if the renormalisation scale, $\mu = x_\mu \times E_{cm}$, is modified — in this case it is actually fitted to the data — from its canonical value of $x_\mu = 1$. Resummation of leading and next-to-leading logarithms, if matched with the NLO calculations, are superior to pure NLO predictions, however still seem to require additional optimisation of the renormalisation scale.

Determinations of $\alpha_s$ from hadronic event shapes and jet rates measured at LEP, taking into account hadronisation effects as predicted from model calculations, are sum-marised on **Slide 16** [47]. Experimental uncertainties are given by the innermost error bars; the total errors also include theoretical (mostly renormalisation scale) as well as hadronisation uncertainties. Results for c.m. energies below the $Z^0$ resonance, i.e. $E_{cm} < 91$ GeV, are from radiative events measured at LEP [48] and from a reanalysis of JADE data [49] based on identical methods and observables as utilised at LEP.

## 10  Renormalisation Scale Dependence of QCD Predictions

In most QCD studies at LEP, the dominating error comes from theoretical uncertainties. These are due to the fact that predictions exist only to finite, truncated perturbative order. Higher order uncertainties are usually quantified by studying the renormalisation scale dependence of the respective predictions. There are no generally accepted procedures to define the range and central values of scale changes and — in general — of higher order uncertainties to different observables. Respective theoretical errors are therefore to be regarded as estimates, and in fact in many cases constitute lower limits rather than complete error determinations.

The size of renormalisation scale uncertainties for LO, NLO and NNLO QCD predictions are demonstrated on **Slide 17**, for the example of the normalised hadronic partial decay width of the $Z^0$ boson, $R_Z$. Theoretical predictions of $R_Z$ are available up to complete NNLO perturbation theory [22, 50]. The scale dependence of $\alpha_s$, calculated from the LEP value of $R_Z = 20.768 \pm 0.0024$, is displayed as a function of the scale factor $x_\mu = \mu/E_{cm}$. In NLO, the strong scale dependence of $\alpha_s$ is partly compensated by the scale dependent QCD coefficient in front of the $(\alpha_s/\pi)^2$ term, leading to a distinct minimum of the curve near $x_\mu = 0.7$. Theoretical proposals to fix the scale ambiguity in NLO QCD [51] are given by the symbols displayed in the Figure. In NNLO, the scale dependence is much reduced, leading to both a minimum and a maximum of the functional dependence of $\alpha_s$ from $x_\mu$. Taking the region between the minimum and the maximum as an estimate of the scale uncertainty, with the central value quoted at $x_\mu = 1$, results in a scale error of $^{+0.003}_{-0.001}$ [21]. Note, however, that when accepting a scale range beyond the one chosen here, the error in $\alpha_s$ rapidly inflates in both directions, even in NNLO QCD.

Renormalisation scale uncertainties in $\alpha_s$ from hadronic event shapes, which are known only to complete NLO, are further studied and quantfied in a study [45] summarised on **Slide 18**. To the left, the scale dependence of $\alpha_s(M_{Z^0})$ and of $\chi^2$ of fits to various event shape observables measured at LEP-I is shown, demonstrating that different observables exhibit different but distinct sensitivities to the renormalisation scale. Experimental optimisation of the renormalisation scale to describe measured distributions best (c.f. **Slide 15**), leads, in general, to a good agreement of the values of $\alpha_s(M_{Z^0})$ from different observables in NLO QCD, see the centre and the right hand graphs on **Slide 18**.

## 11  Power Corrections

Analytic approaches to approximate nonperturbative hadronisation effects on event shape observables lead to power corrections [17] to the respective perturbative QCD predictions. These corrections, depending on a parameter $\alpha_0$ which is predicted to be universal for all observables, provide a consistent description of measurements of event shape observables in a wide range of c.m. energies, see the left hand graph of **Slide 19** for mean values of the C parameter [52] and the right hand graph for differential distributions of the jet broadening parameter $B_T$ [53]. A summary of all measurements of $\alpha_s(M_{Z^0})$ and of $\alpha_0$ [54], based on mean values of hadronic event shapes at LEP and in deep inelastic scattering at HERA, and on studies of differential shape distributions at LEP, using NLO QCD plus power corrections, is shown on **Slide 20**.

## 12 Gluon Splittings into Heavy Quarks

A summary [55] of measurements [56] of gluon splitting into heavy quarks, $g_{b\bar{b}}$ and $g_{c\bar{c}}$, and of the relative rate of 4-jet events with two $b\bar{b}$ quark pairs in the final state, $g_{4b}$, is given on **Slide 21**, demonstrating that the measurements are in broad agreement with the QCD predictions.

## 13 Running b-quark Mass and Flavour Independence of $\alpha_s$

Based on NLO QCD calculations of 3-jet event rates for massive quarks [57], determinations of the running b-quark mass were performed at LEP and at the SLC [58], analysing the ratio of 3-jet event production rates in tagged $b\bar{b}$ and in light quark events. The left hand graph on **Slide 22** shows the dependence of this ratio on the jet resolution parameter $y_{\rm cut}$, while the right hand graph summarises current results on $m_b(M_{Z^0})$ and on $m_b$ determined from $\Upsilon$ decays at lower energy scales. The results are in good agreement with the QCD expectations of a running b-quark mass. Likewise, from similar studies of 3-jet event production rates in tagged heavy quark and light quark events, the flavour independence of $\alpha_s$ was proven, to accuracies of a few per-cent, at LEP; see **Slide 23** [55, 59].

## 14 Two Photon Physics: Structure Function $F_2^\gamma$ and Scaling Violations

Extensive studies of two-photon scattering processes leading to hadronic final states have been performed at LEP. On **Slide 24**, measurements [60] of the photon structure function $F_2^\gamma$ as a function of the Bjorken scale $x$ are summarised [61], demonstrating, for the first time, a possible increase of $F_2$ at small $x$. Scaling violations of $F_2$ as a function of $Q^2$ are visible in the compilation of data from LEP and from previous $e^+e^-$ experiments [61].

## 15 Two Photon Physics: Heavy Quark Production

The production of heavy quark final states in two-photon scattering processes, measured in a broad range of $e^+e^-$ centre of mass energies at LEP and at previous experiments, is summarised on **Slide 26** [62]. While the production of charm quark-antiquark pairs [63] is in good agreement with NLO QCD predictions, the production of bottom quark final states is observed at roughly 4 times the predicted rate, see [62, 64], leaving a puzzle to be solved by future measurements and calculations.

## 16 Summary and Conclusion

The successful running of the LEP electron-positron collider at CERN has led to a significant increase of knowledge about hadron production and the dynamics of quarks and gluons at high energies, as summarised on **Slide 27**. Precise determinations of $\alpha_s$ at the smallest and the largest c.m. energies available to date, experimental confirmation of asymptotic freedom and of the gluon self coupling, detailed studies of differences between quark and gluon jets, verification of the running b-quark mass and of the flavour independence of $\alpha_s$, deeper understanding of power corrections and of hadronisation models to describe the nonperturbative hadronisation domain, and detailed studies of hadronic systems in two-photon scattering processes were summarised in this report, proving QCD as a consistent theory which accurately describes the phenomenology of the Strong Interaction. Future developments in this field will require further NNLO QCD calculations and predictions, especially for jet rates and shape observables, in order to assess and minimise the current QCD uncertainties in a more reliable way.

After LEP was finally shut down a month after this report was presented, further results obtained from LEP data were published, and more analyses are expected to be completed in the future. These will especially comprise results from the LEP QCD Working Group, which currently prepares to combine results from the four LEP experiments in a consistent manner, with the aim to increase the overall significance and to obtain a consistent treatment of systematic, experimental and theoretical uncertainties.

Le LEP est mort — vive le LHC [65]! And of course, a future $e^+e^-$ Linear Collider!

S. Bethke
MPI für Physik
Munich

> 200 journal pub.'s on hadronic physics
> 30 journal publications on $2\gamma$ physics sorry: today, no...
– BE correlations
– Intermittency
– inclus. particle ID
– ... and lots of other interesting studies

Many thanks to:
S. Banerjee
O. Biebel,
K. Hamacher
R. Jones
D. Wicke
S. Söldner-Rembold
H. Stenzel
The LEP QCD WG
Funding Agencies
CERN
SL-Division
all who helped to make LEP a success

# An attempt to summarize 11 years of

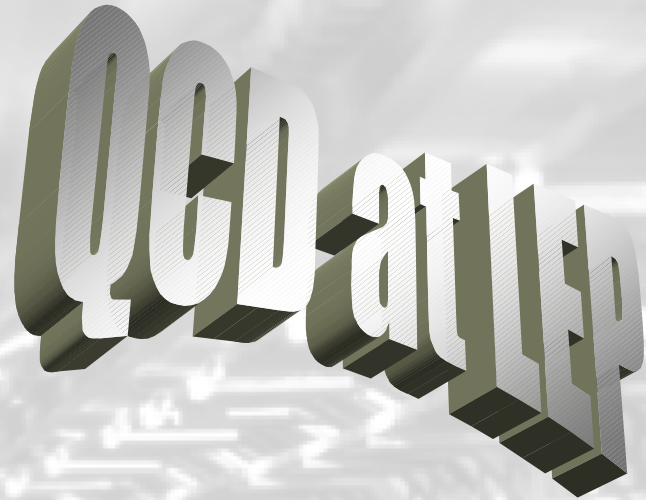

QCD at LEP

- status of QCD before and at the end of LEP
- hadronic events, event shapes, hadronisation
- measurements of $\alpha_s$
- tests of asymptotic freedom
- non-Abelian gauge structure of QCD
- differences between quark- and gluon-jets
- gluon coherence; local parton-hadron duality
- power corrections
- scaling violations
- 2-photon physics

# Anatomy of hadronic events in $e^+e^-$ annihilation

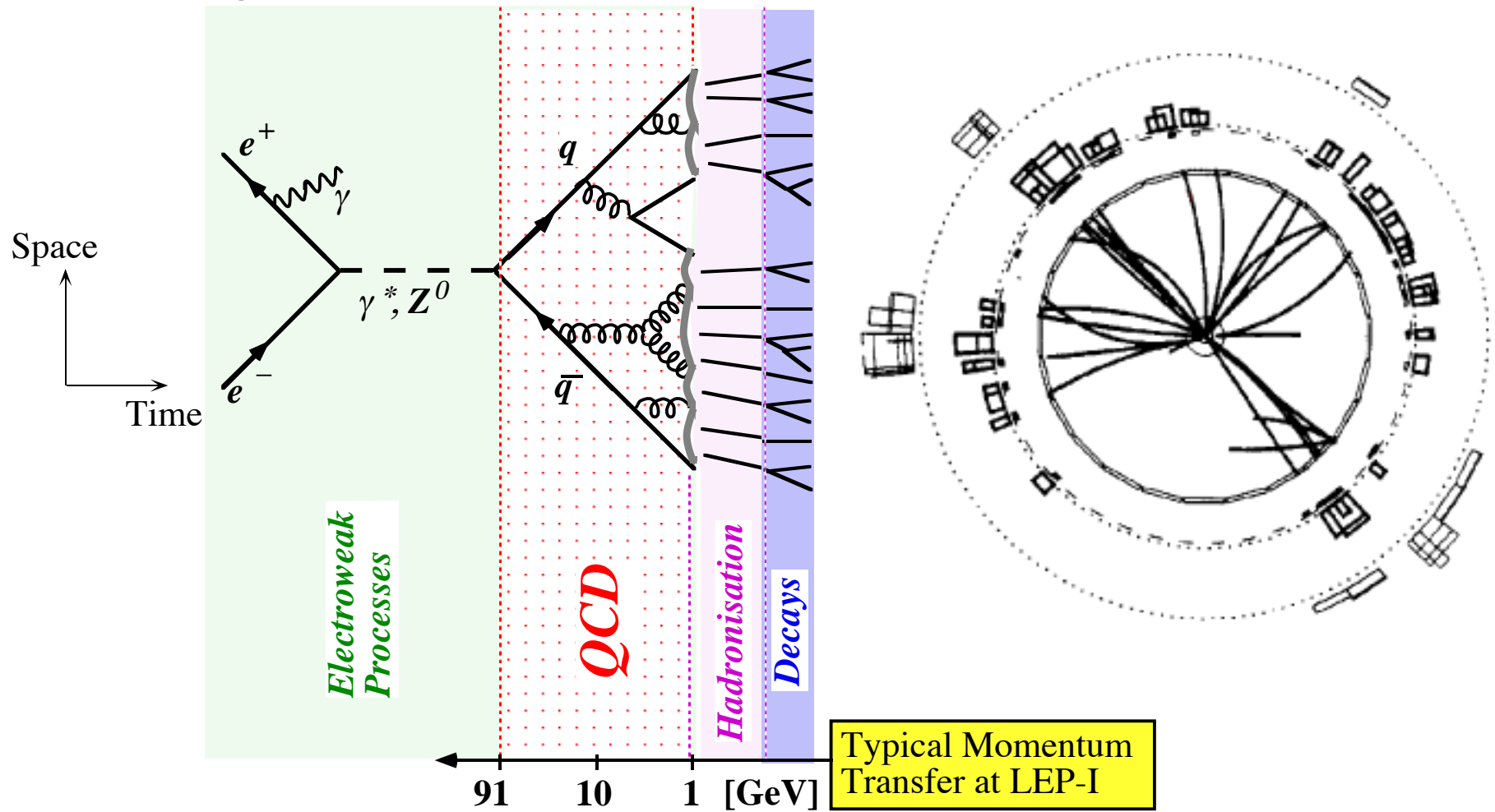

- QCD: shower development calculated in perturbation theory (fixed order; (N)LLA)
- Hadronisation: phenomenological models of string-, cluster- or dipole fragmentation
- Decays: randomized according to experimental decay tables

# Hadronic event recorded at 205.4 GeV c.m.

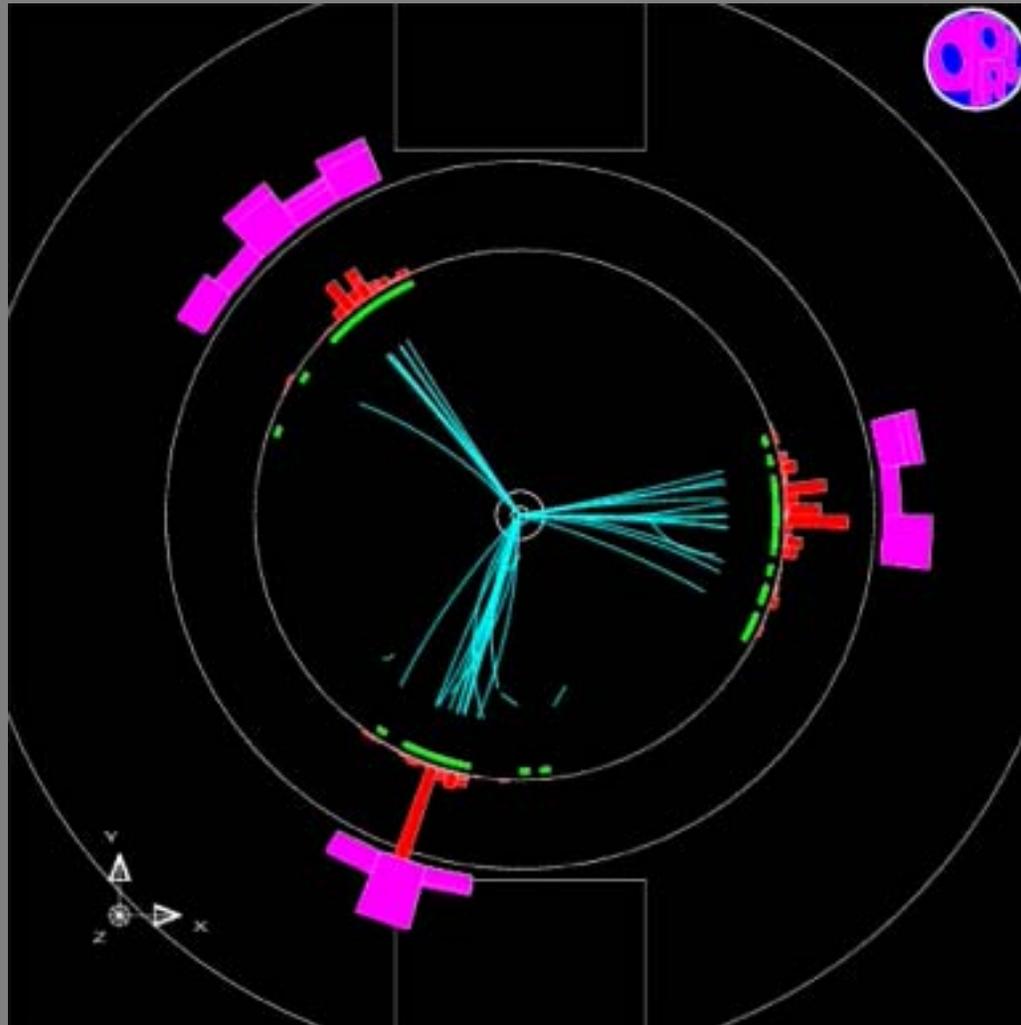

# Improvements at LEP:

- **detectors:** hermetic, homogenous, HCAL's, Si-μ-vertex, better double-track-, energy- and spacial resolutions, particle-ID..

- **data:** high statistics, low background (LEP-I), higher energies (jet collimation), precise c.m. energy, precise luminosity

- **tools:** new observables (e.g. jet broadening $B_T, B_W$), new jet finders (Durham, Cambridge), heavy flavour tagging, improved MC models, neural nets, ...

- **methods:** assessment of theoretical uncertainties, simultaneous analysis of many observables, gluon-jet (anti-)tagging, quark flavour tagging ...

- **theory:** MC integration of ERT NLO matrix elements, resummation, power corrections, NNLO for $\sigma_{had}$, $\Gamma_{had}$ and $R_{tau}$, NLO for 4-jets, NLO for massive quarks, ...

- **plus:** the great experience and knowledge from the PETRA, PEP and TRISTAN experiments ! (and some from SLC, too...)

# $\alpha_s$ in $e^+e^-$ annihilations

## 1989

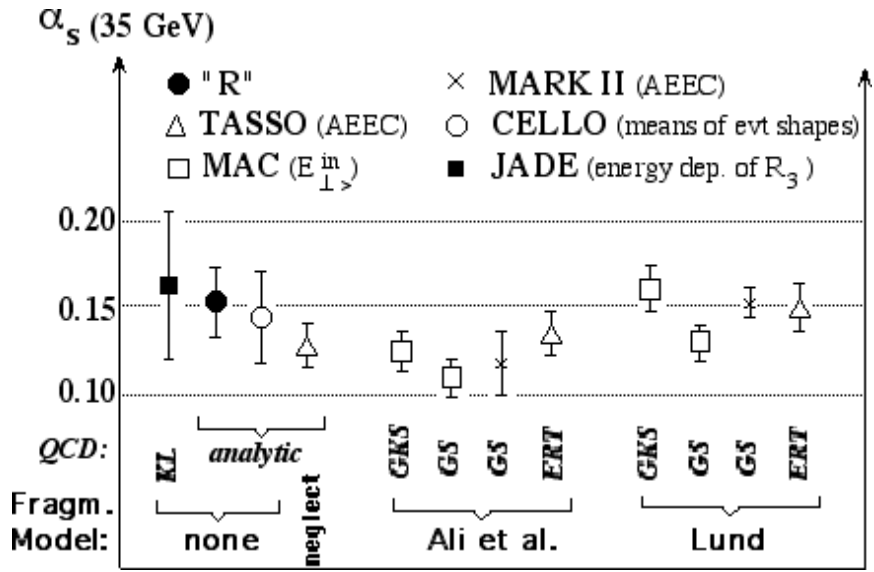

## 2000

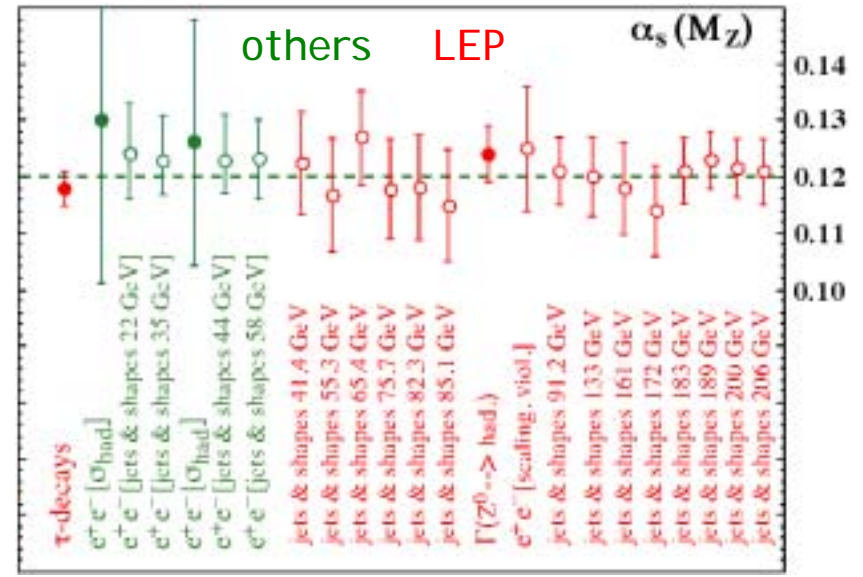

$\alpha_s(35\text{ GeV}) = 0.14 \pm 0.02$ (NLO)

$\Rightarrow \alpha_s(M_Z) = 0.119 \pm 0.016$ (NLO)

$\alpha_s(M_Z) = 0.121 \pm 0.005$ (res. NLO)

$\alpha_s(M_Z) = 0.120 \pm 0.003$ (NNLO)

# World summary of $\alpha_s$

## 1989

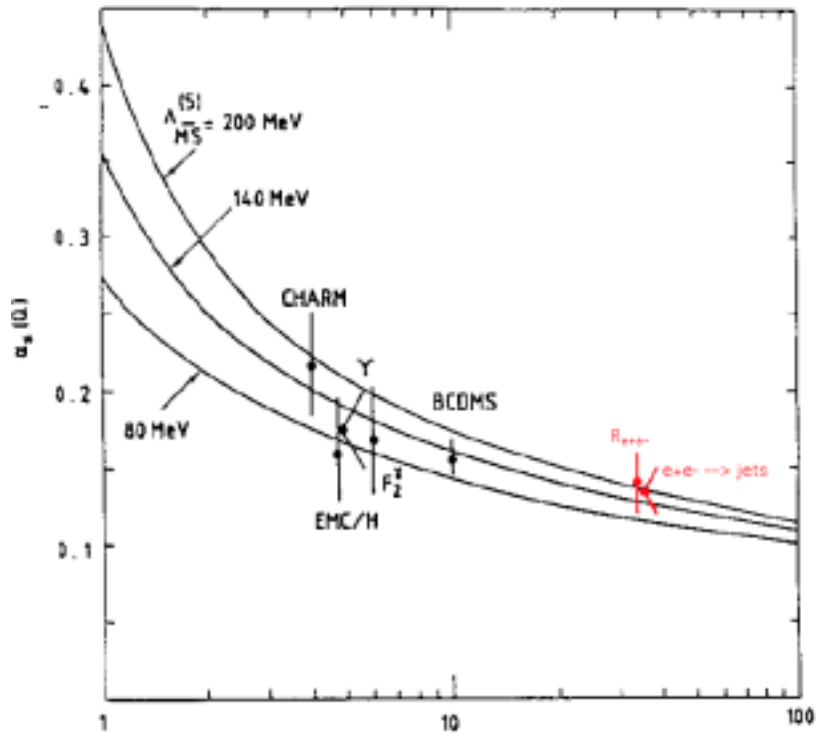

$$\alpha_S(M_Z) = 0.110^{+0.006}_{-0.008} \text{ (NLO)}$$

G. Altarelli, Ann. Rev. Nucl. Part. Sci. 39, 1989

## 2000

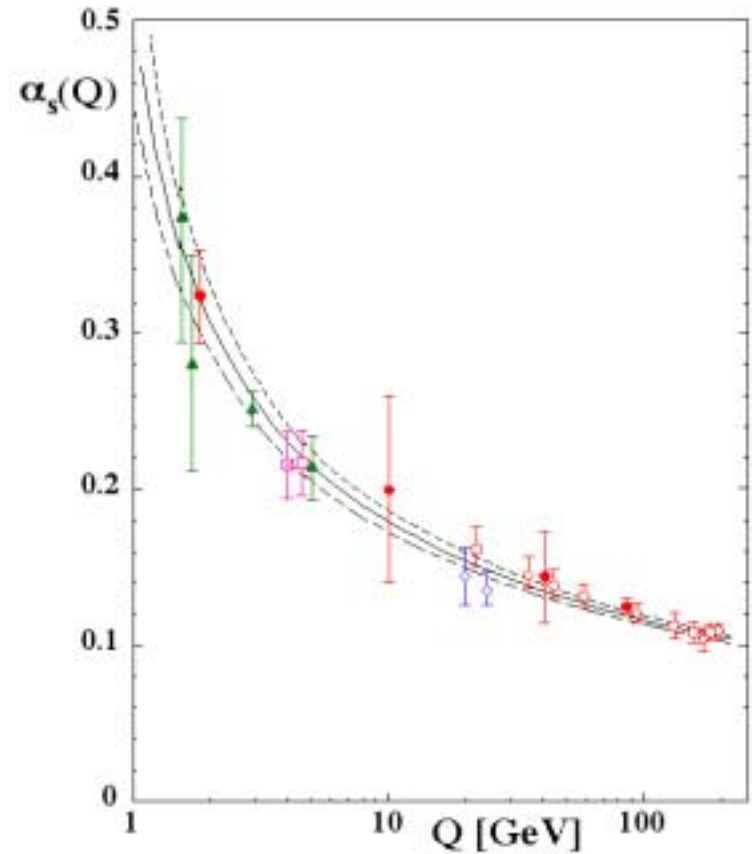

$$\alpha_S(M_Z) = 0.1184 \pm 0.0031 \text{ (NNLO)}$$

S. B. , J. Phys. G 26, 2000

# Asymptotic Freedom (running $\alpha_s$)

energy dependence of 3-jet production rates ($R_3$):

$$R_3 = C_1(y_{cut}) \cdot \alpha_s(\mu) + C_2(y_{cut}) \cdot \alpha_s^2(\mu)$$

JADE Jet finder:
small and (almost) energy independent
hadronisation corrections:

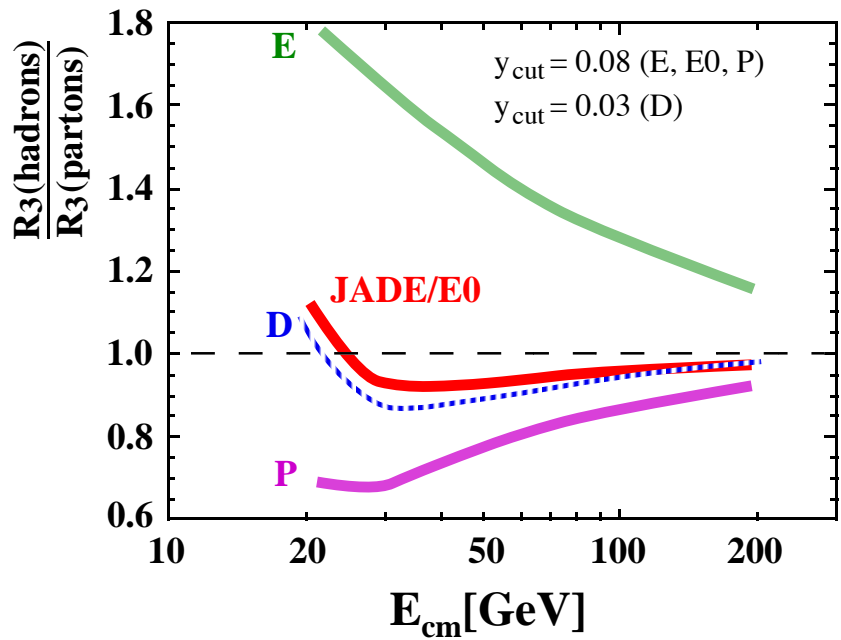

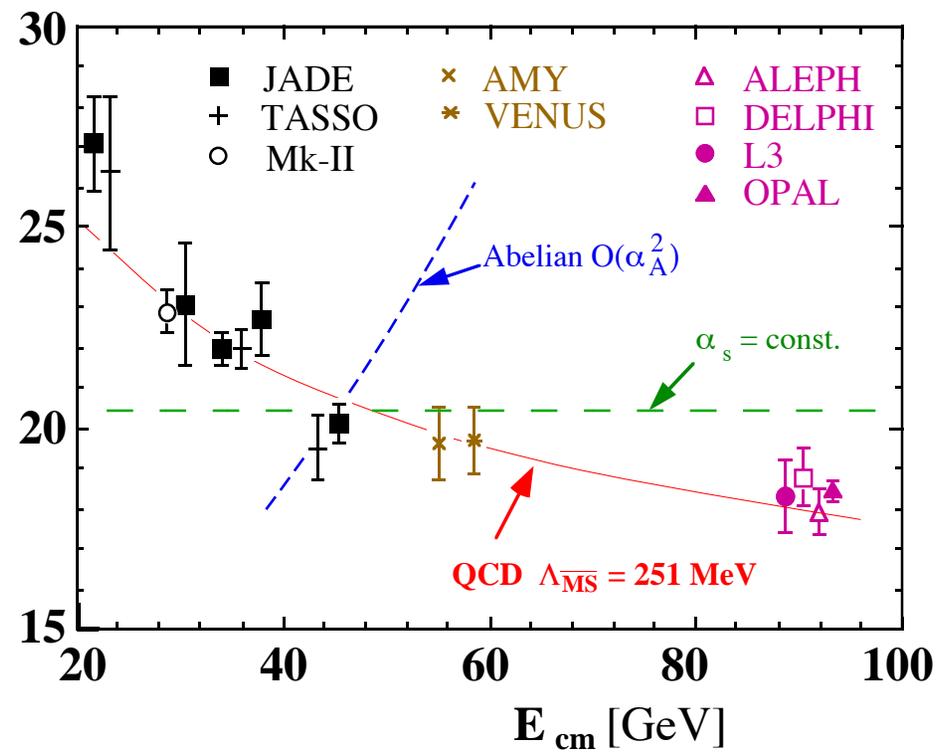

# Asymptotic Freedom from jet rates

$$R_3 \equiv \frac{\sigma_{3-\text{jet}}}{\sigma_{\text{tot}}} \propto \alpha_s(E_{cm}) \propto \frac{1}{\ln E_{cm}}$$

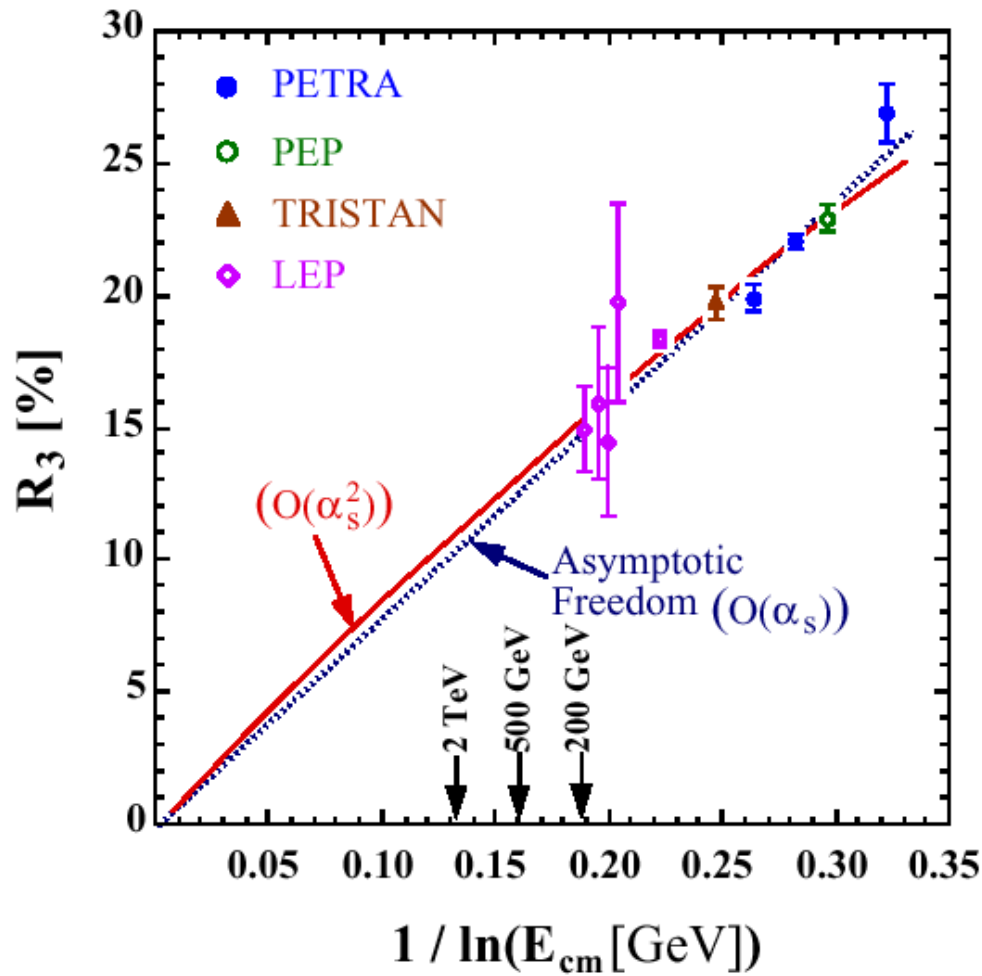

# Non-Abelian gauge structure from 4-jet events

Bengtson-Zerwas angle between energy-ordered jet axes

($E_1 \geq E_2 \geq E_3 \geq E_4$)

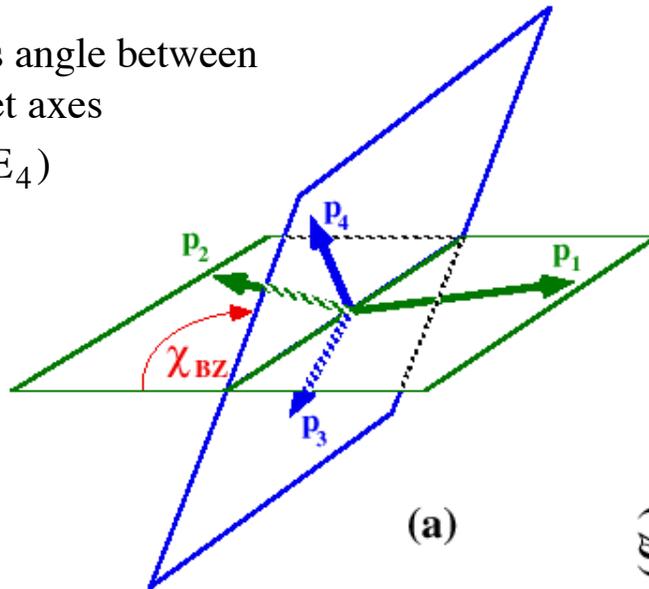

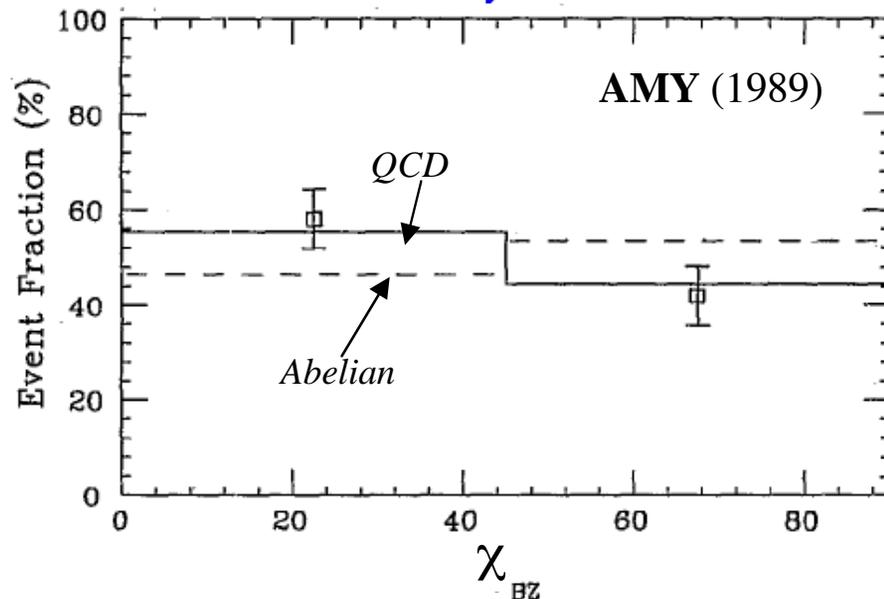

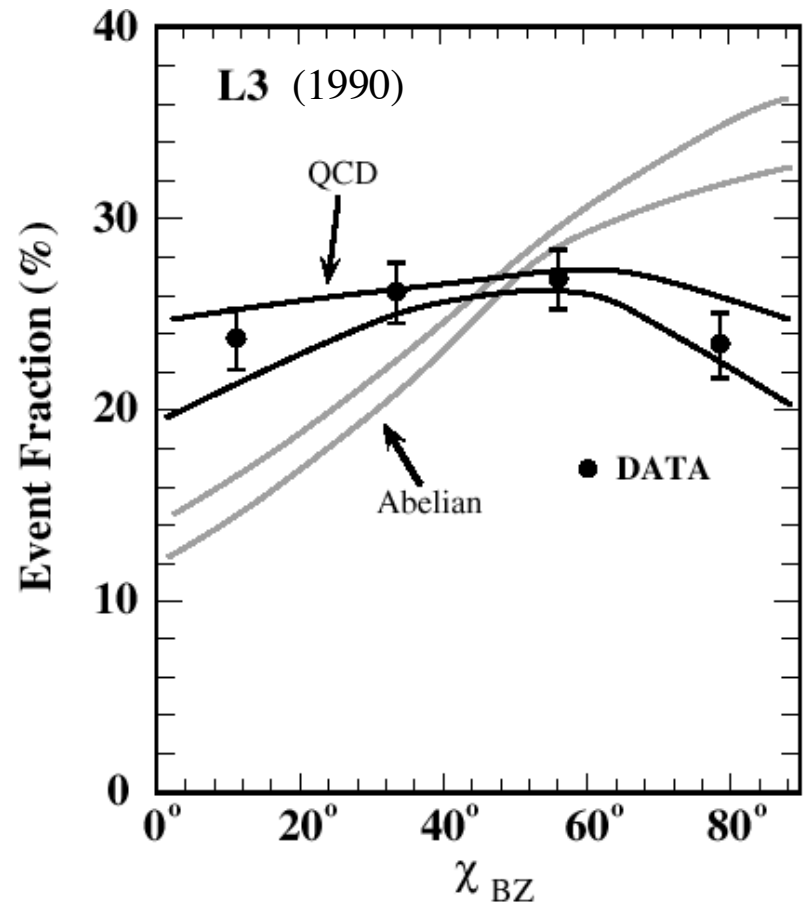

# Non-Abelian gauge structure from 4-jet events

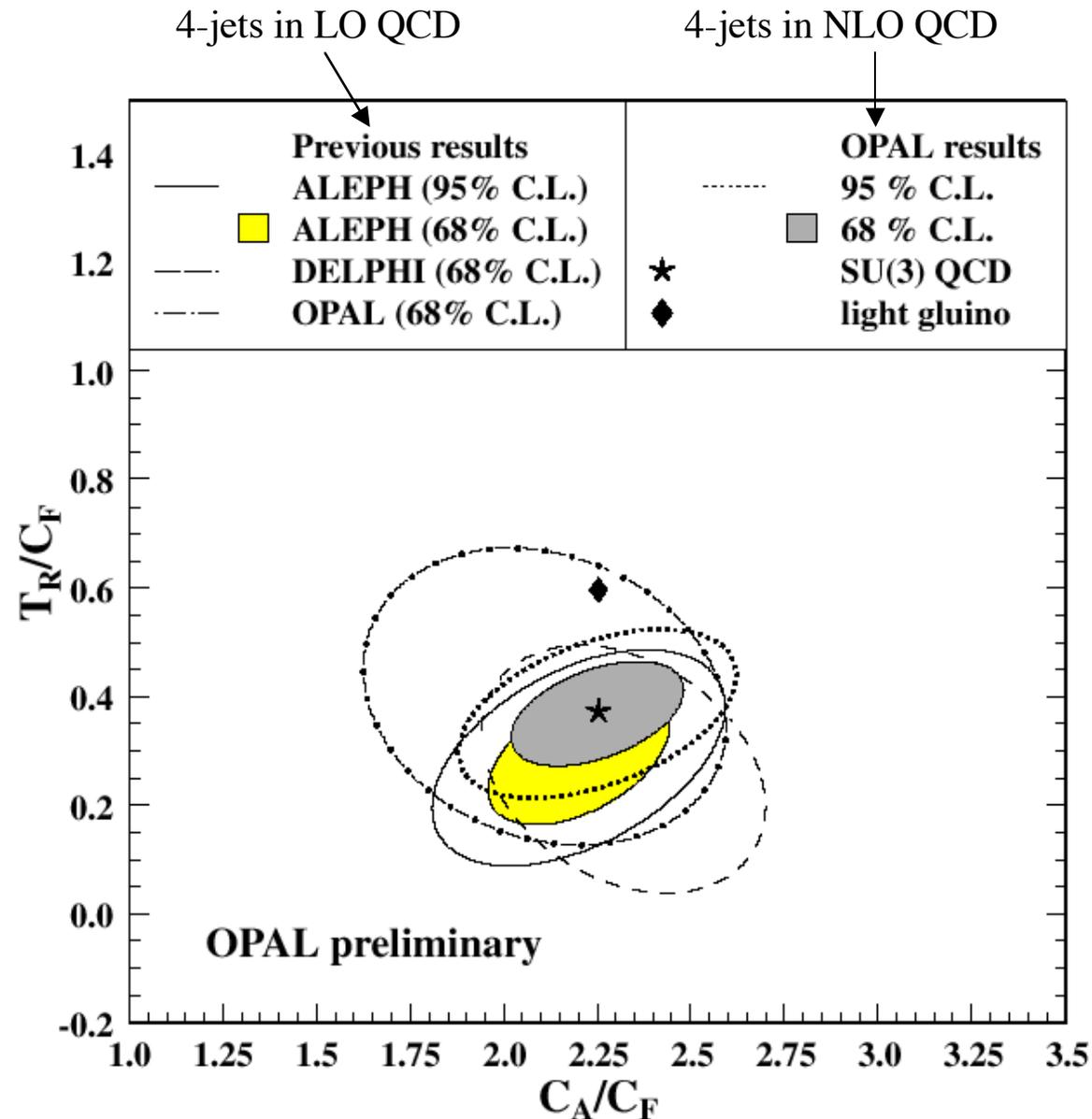

# Differences between q- and g-jets

"naive" QCD-expectation:
(i.e. at ∞ energies, in leading order perturbation theory)

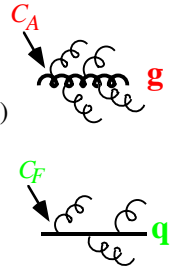

- Gluonjets are 'broader' than Quarkjets
- particles in Gluonjets are 'softer' (i.e. they are less energetic)
- $\dfrac{N_{had}(\text{g-jet})}{N_{had}(\text{q-jet})} \approx \dfrac{C_A}{C_F} = \dfrac{3}{4/3} = \dfrac{9}{4}$

JADE 1982: Energy ordered 3-jet events ($E_1 \geq E_2 \geq E_3$)

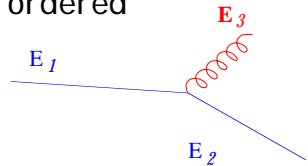

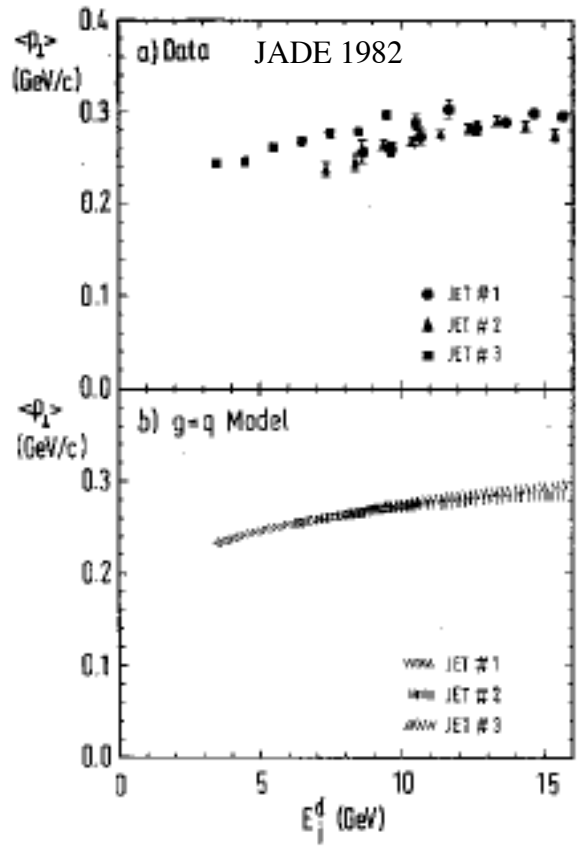

OPAL 1991: antitagging of g-jets in symmetric 3-jet events

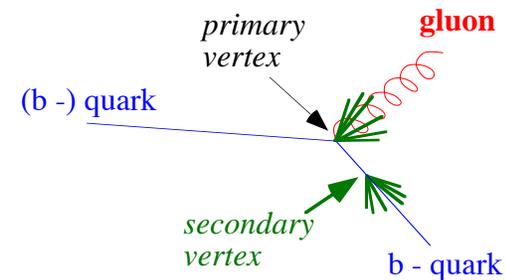

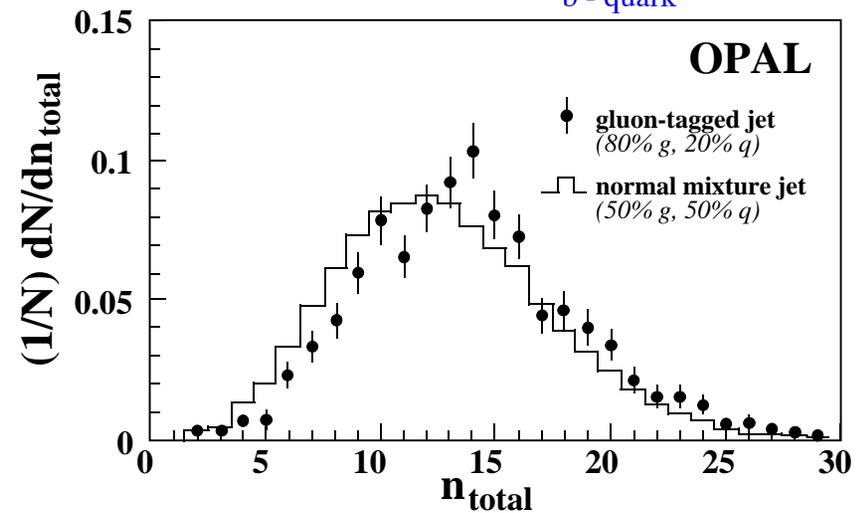

# Differences between q- and g-jets

g recoiling vs. $q\bar{q}$ : direct comparison with QCD prediction

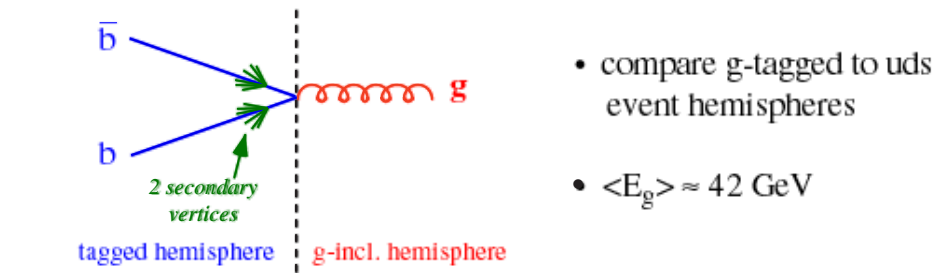

tagged hemisphere | g-incl. hemisphere

- compare g-tagged to uds event hemispheres
- $\langle E_g \rangle \approx 42$ GeV

Delphi (2000): determination of $N_{ch}$ for gg-events from symmetric 3-jet events and from multiplicities of average hadronic ($q\bar{q}$) events, based on theoretical predictions of Eden et al.

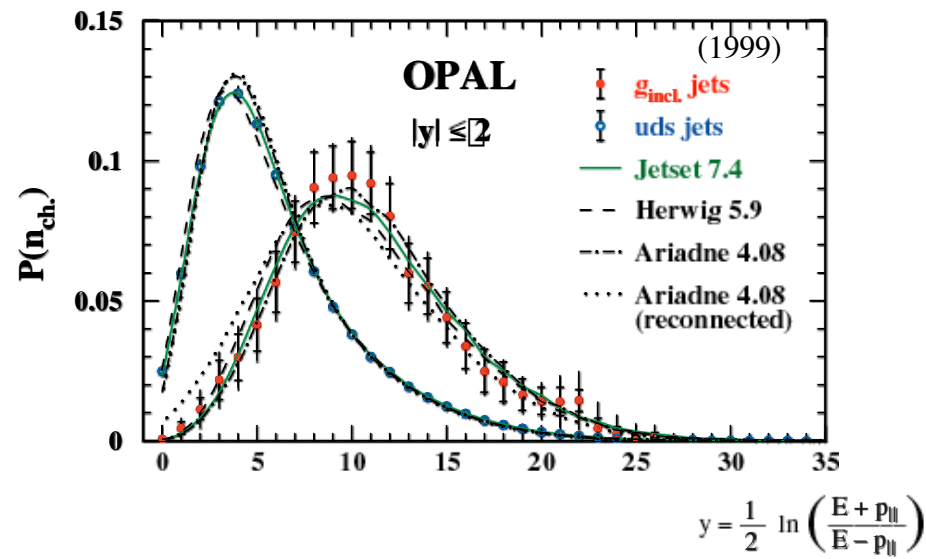

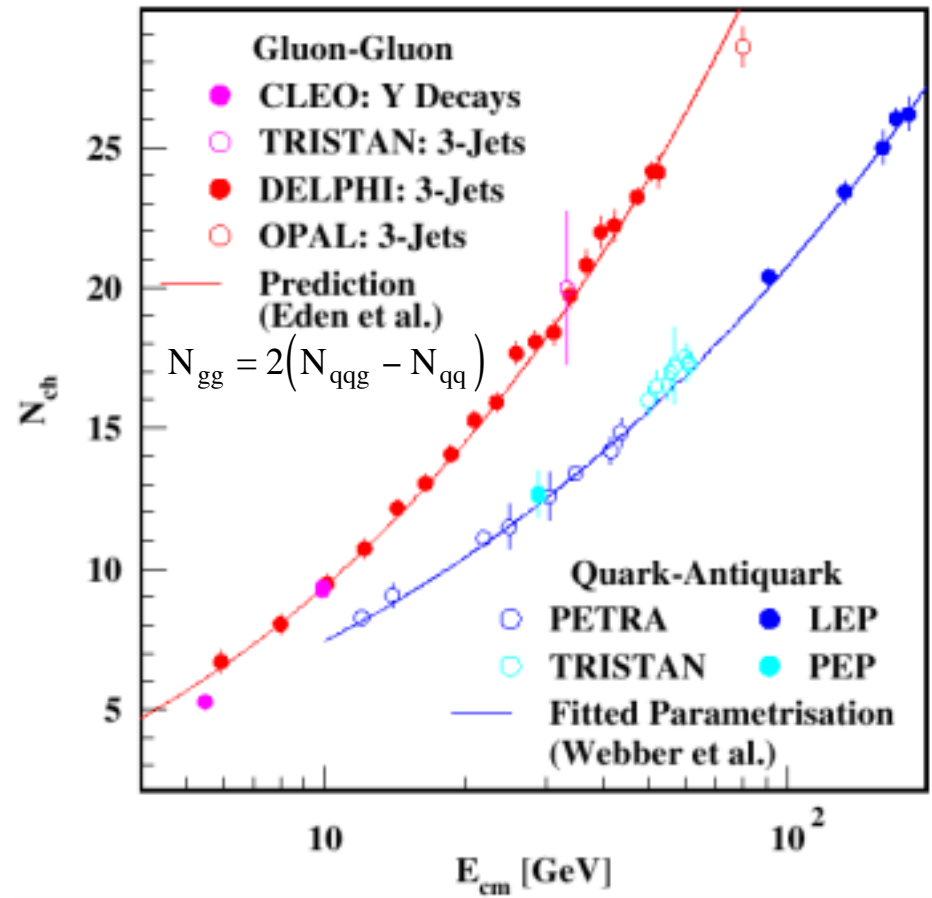

$$N_{gg} = 2(N_{qqg} - N_{qq})$$

$$r_{ch} = \frac{n_{ch}(g_{incl.})}{n_{ch}(uds)} = 1.509 \pm 0.022 \text{ (stat.)} \pm 0.046 \text{ (sys.)} \quad \text{(all y)}$$
$$= 1.815 \pm 0.038 \text{ (stat.)} \pm 0.062 \text{ (sys.)} \quad (|y|<2)$$
$$= 1.87 \pm 0.05 \text{ (stat.)} \pm 0.12 \text{ (sys.)} \quad (|y|<1)$$

Remaining difference to r = 2.25 (asymptotic QCD prediction) due to finite energy effects

$\Rightarrow C_A / C_F = 2.22 \pm 0.11$

# String-effect and hadronisation models

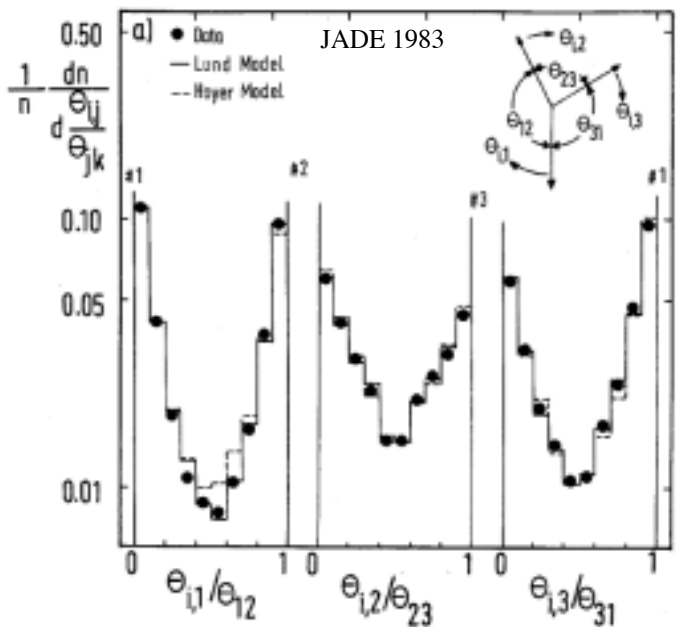

JADE 1983

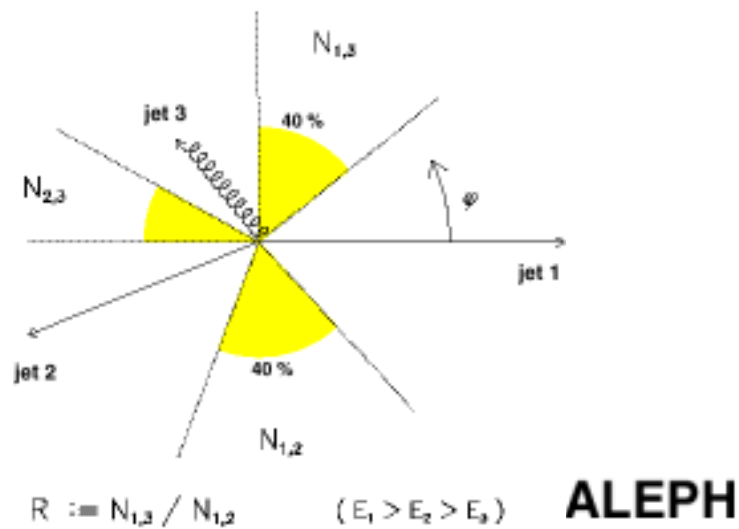

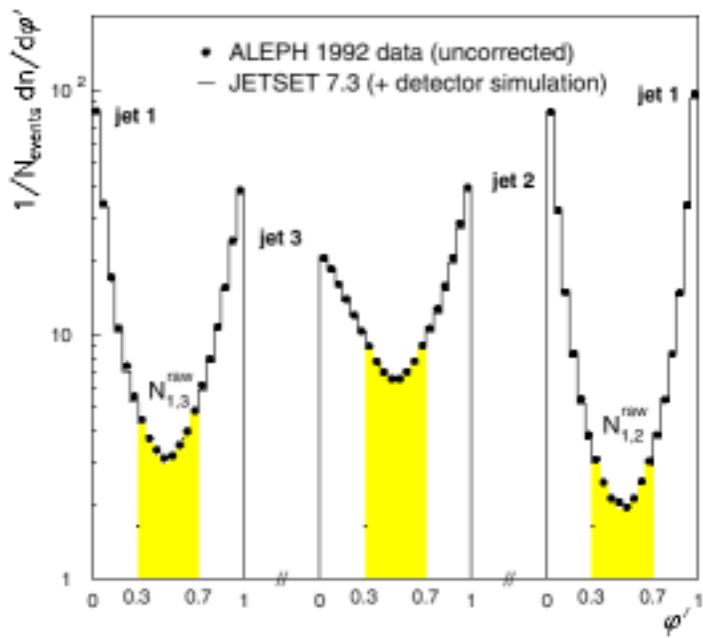

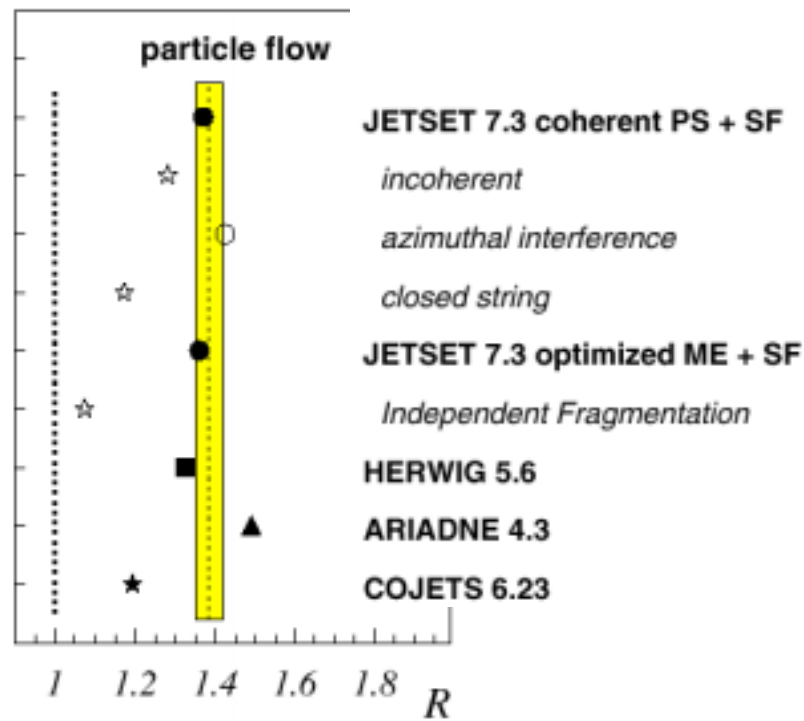

# Jet production and hadronic event shapes

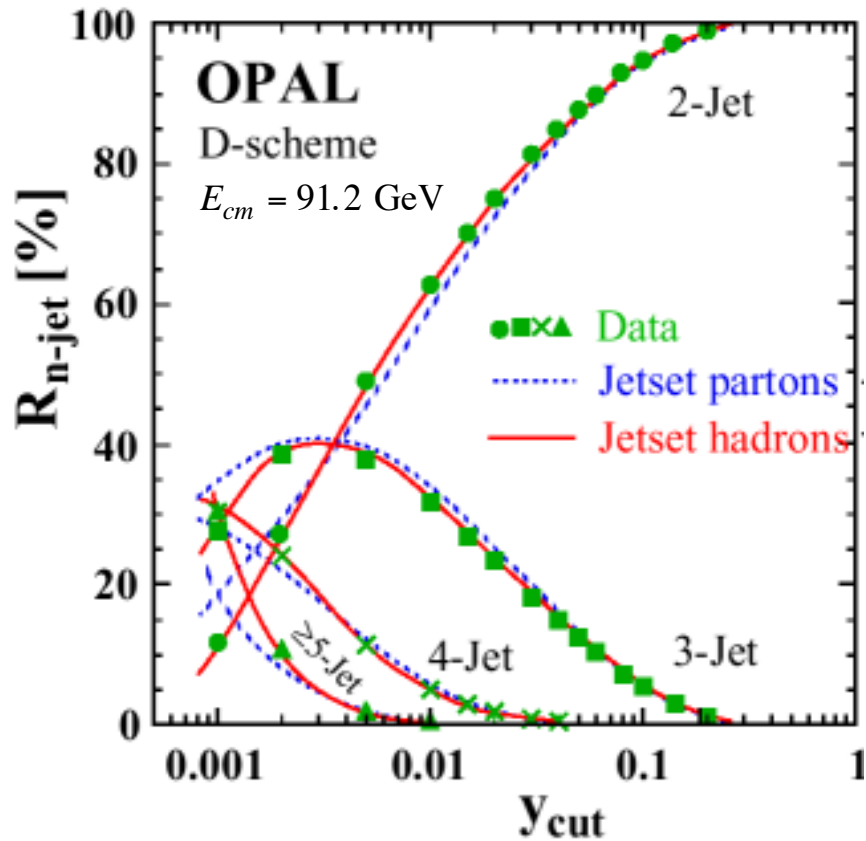
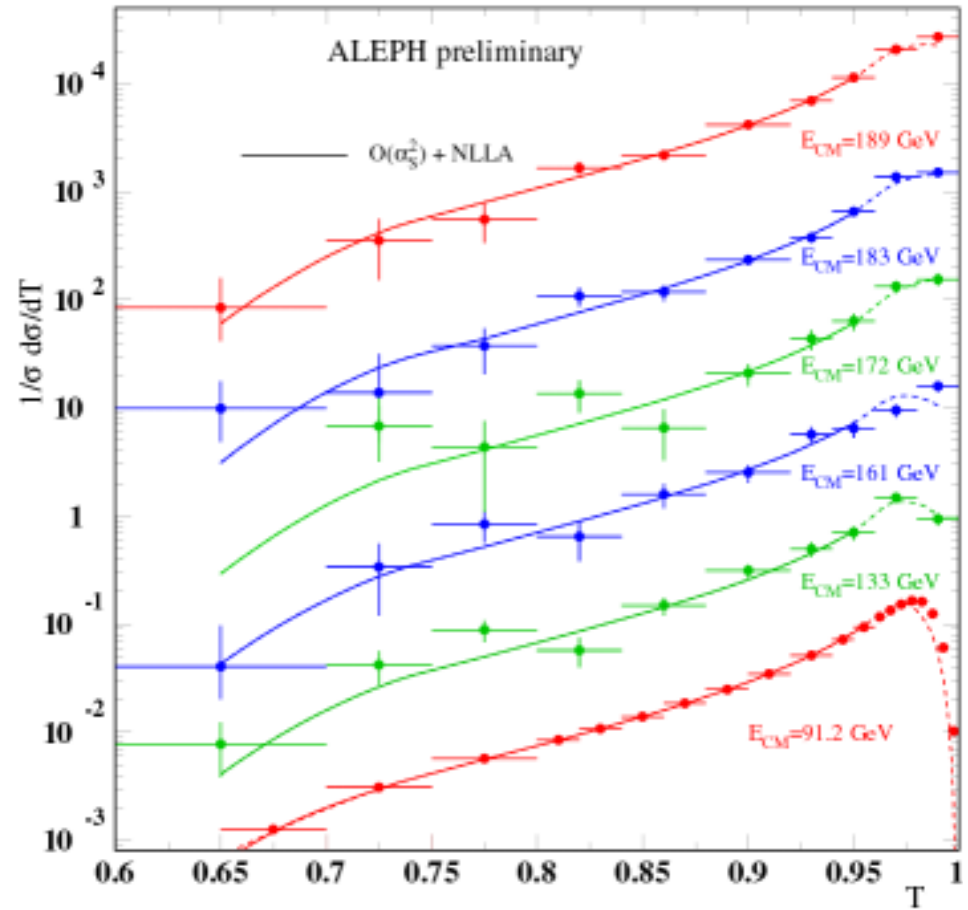

Pert. QCD

- in NLO: $\dfrac{1}{\sigma_0}\dfrac{d\sigma}{dy} = R_1(y)\,\alpha_s(\mu^2) + R_2\!\left(y,\dfrac{\mu^2}{Q^2}\right)\alpha_s^2(\mu^2)$    Ellis, Ross & Terrano (ERT); Kunszt & Nason, Catani & Seymour

- plus resummation of leading and next-to-leading    Catani, Trentadue, Turnock, Webber
  logarithms (NLLA) –> "matching schemes"

# Precision event shapes and pert. QCD

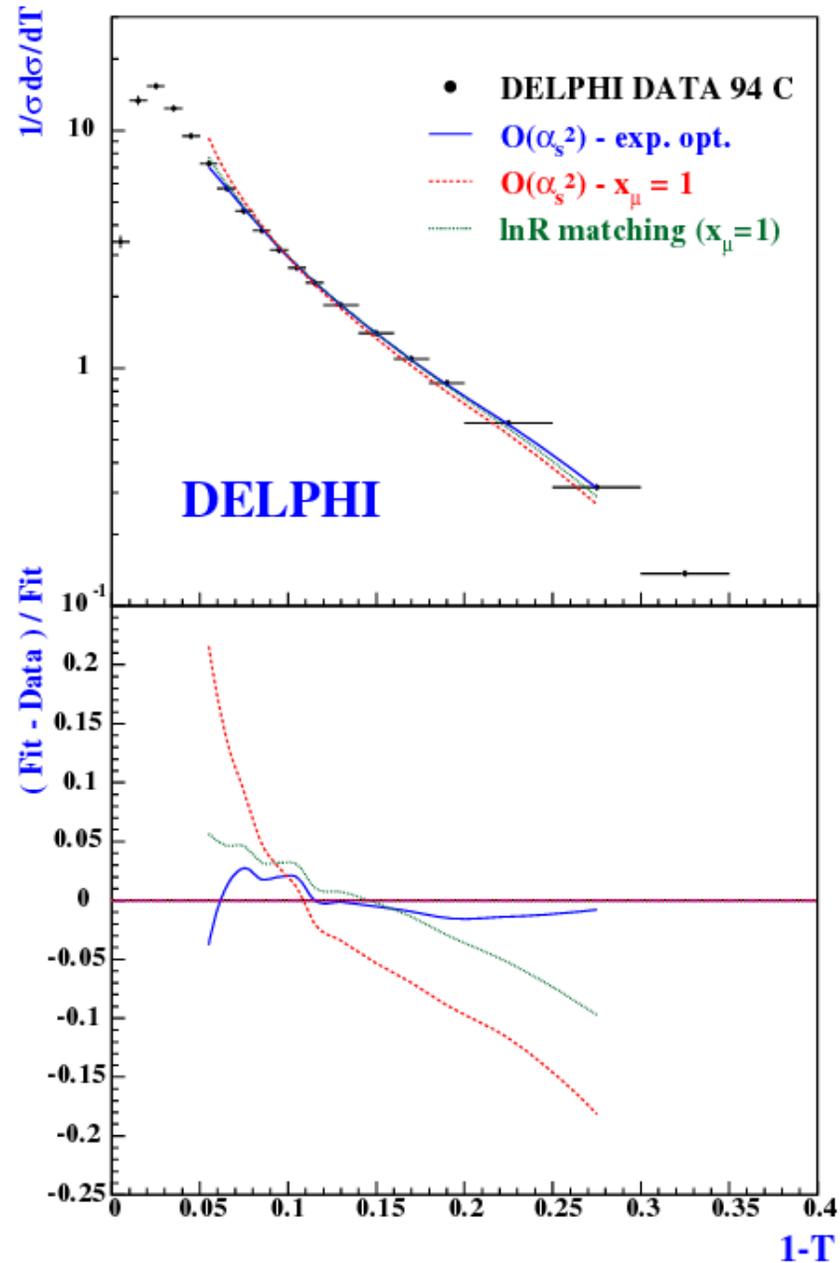

# $\alpha_s$ from event shapes and jet rates

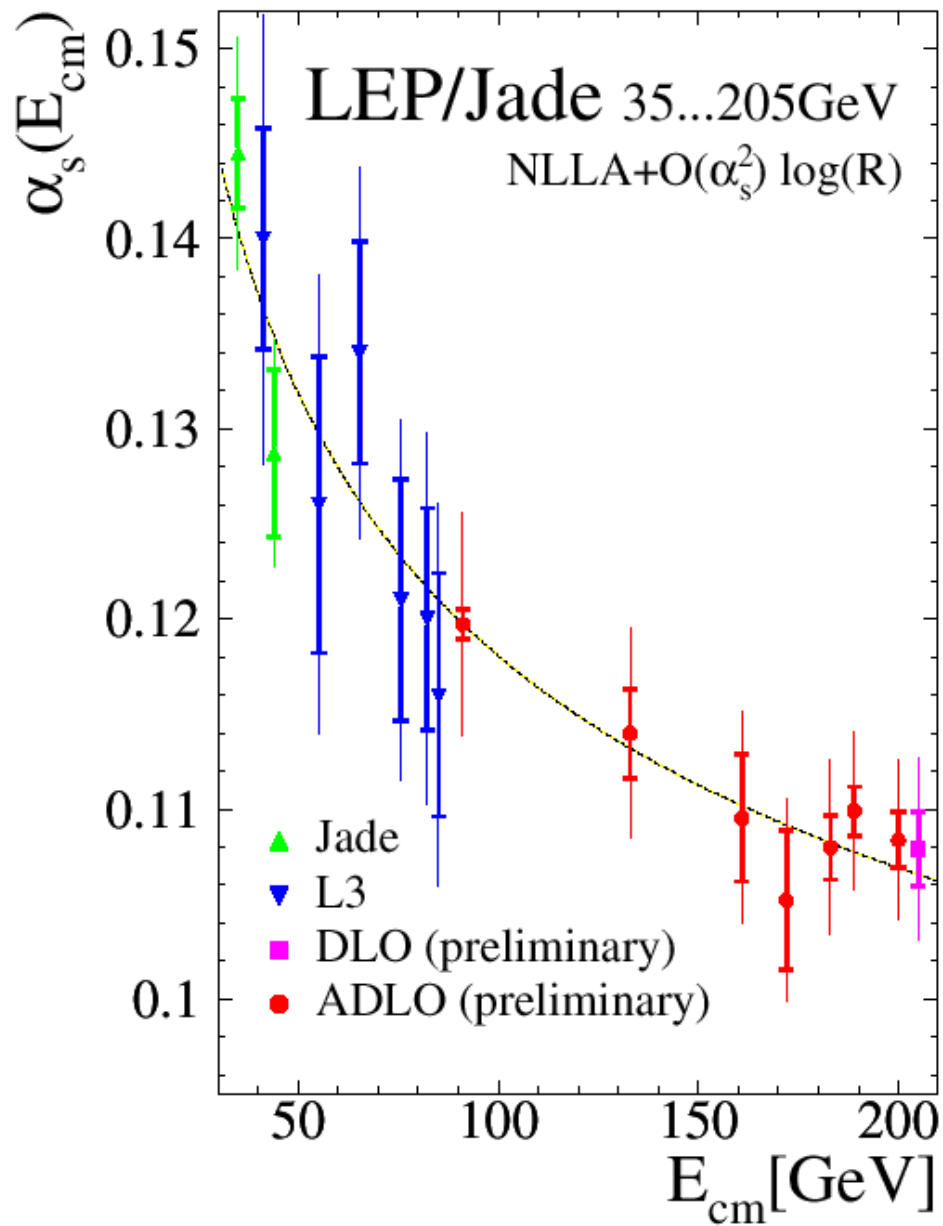

LEP QCD Working Group

# Renormalisation scale dependence

$$R_Z = \frac{\Gamma(Z^0 \to \text{hadrons})}{\Gamma(Z^0 \to \text{leptons})} = 20.768 \pm 0.0024$$

$$R_Z = 19.934 \left[ 1 + 1.045 \frac{\alpha_s(\mu)}{\pi} + 0.94 \left[\frac{\alpha_s(\mu)}{\pi}\right]^2 - 15 \left[\frac{\alpha_s(\mu)}{\pi}\right]^3 \right]$$

Larin, van Ritbergen,, Vermaseren, Chetyrkin, Tarasov, Kühn, Steinhauser, Hoang,......

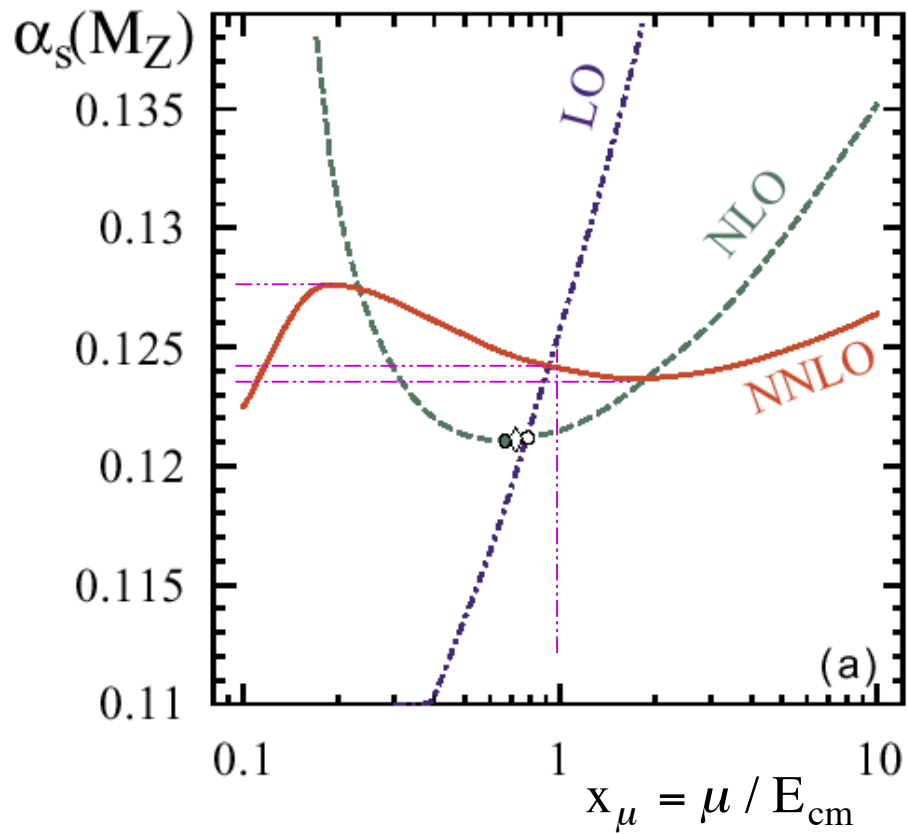

$$\Rightarrow \alpha_s(M_Z) = 0.124 \pm 0.004 \quad (\text{exp.})$$
$$\pm 0.002 \quad (M_H, M_{top})$$
$${}^{+\,0.003}_{-\,0.001} \quad (\text{QCD})$$

# Renormalisation scale dependence in NLO

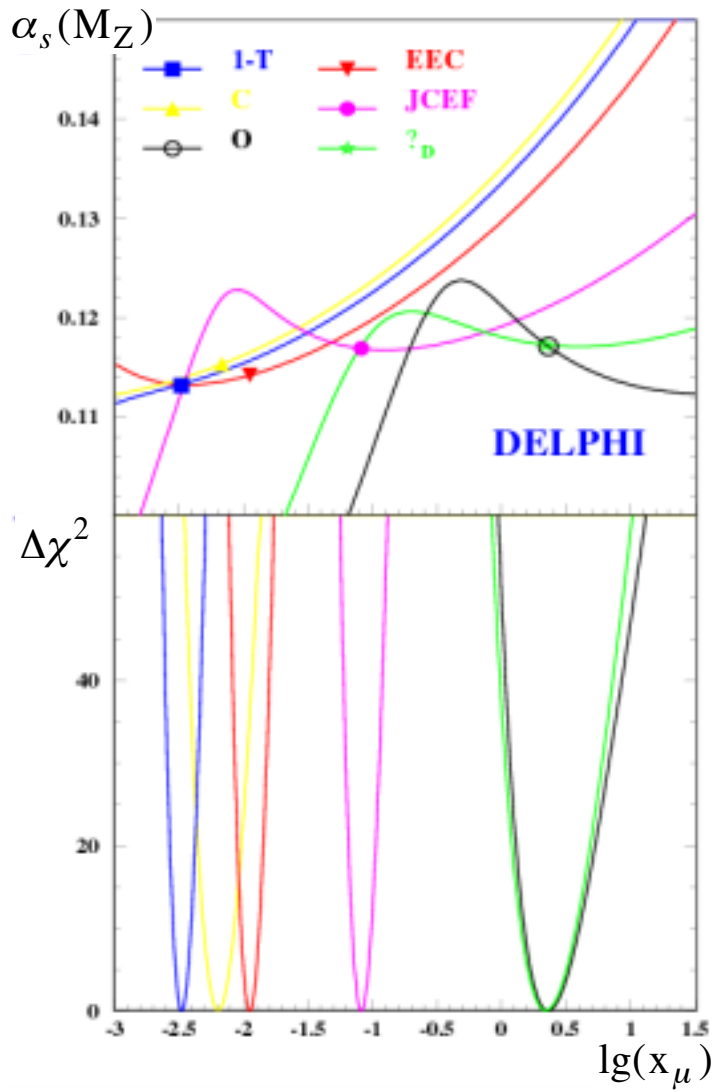
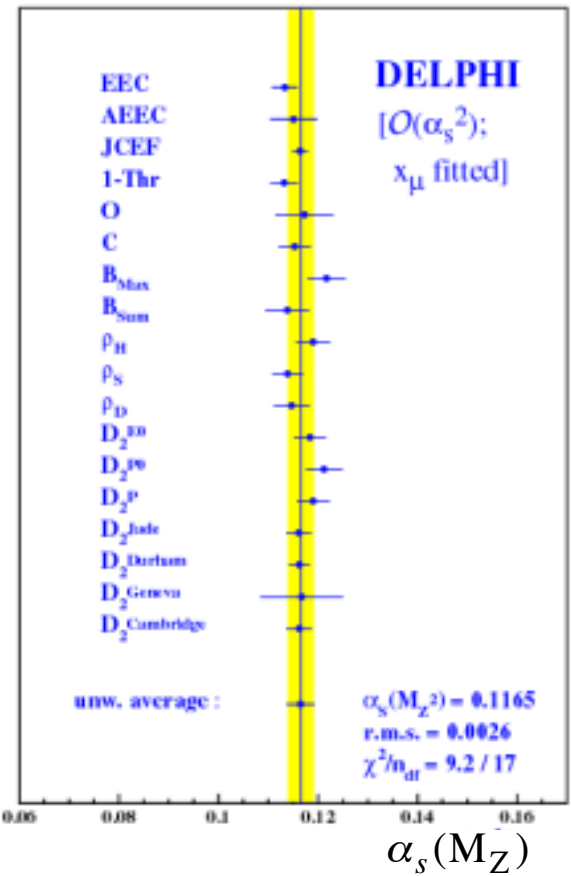
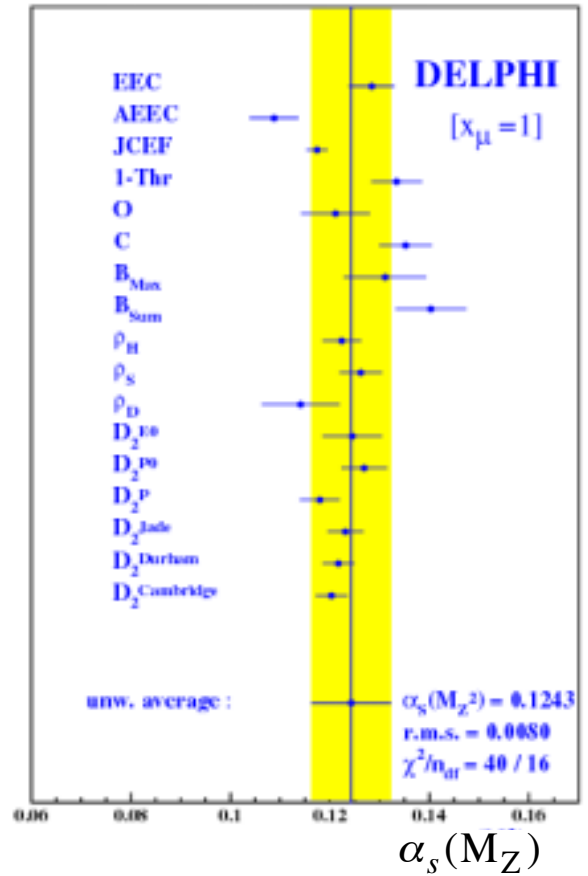

(event shapes and jet rates)

- exp. scale optimisation gives consistent results in NLO
- how to define the corresponding scale uncertainty?

# Power corrections

Analytical approach to approximate nonperturbative hadronisation effects, introducing a universal parameter

$$\alpha_0(\mu_i) = \frac{1}{\mu_i} \int_0^{\mu_i} dk\, \alpha_s(k)$$

→ corrections $\propto 1/Q$, as alternative to hadronisation Models.

*(Dokshitzer, Webber, Marchesini, Catani,…)*

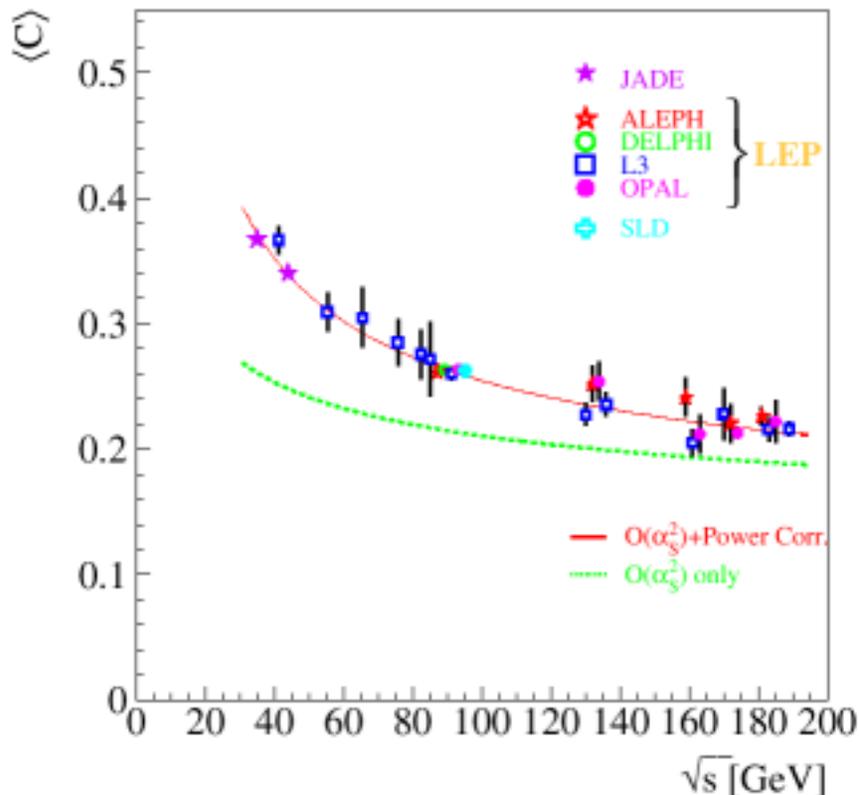
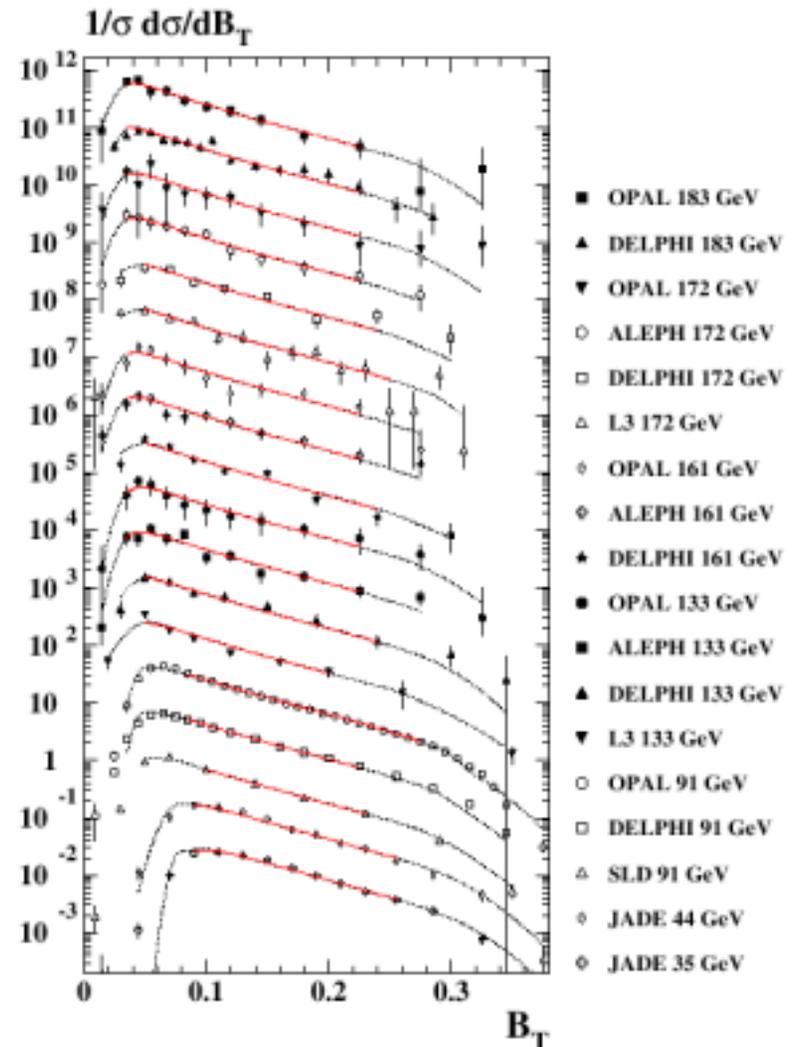

# Power corrections

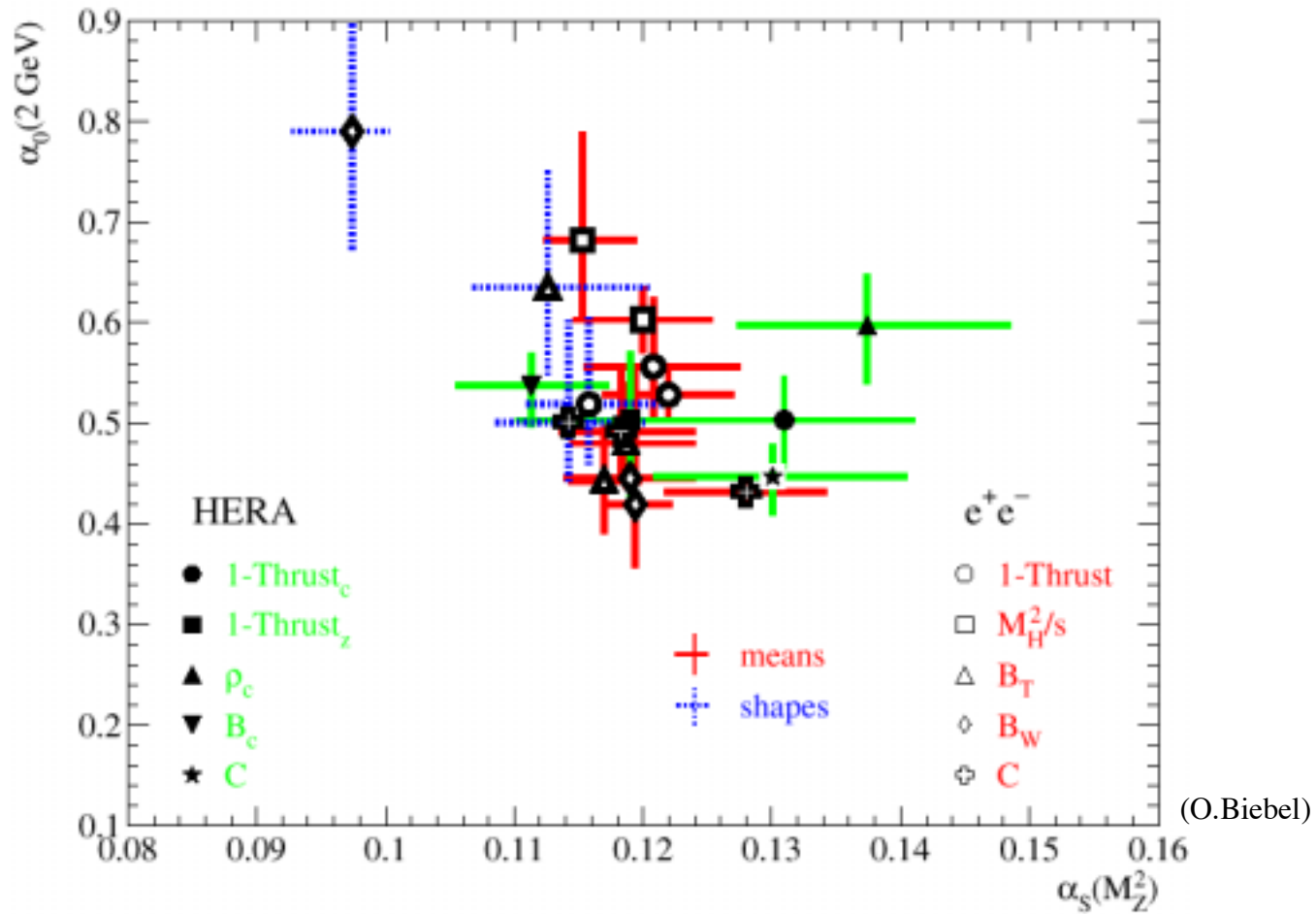

(O.Biebel)

- $\alpha_s(M_z)$ from mean values  ~ 0.120  (similar to analyses using hadronisation models)
- $\alpha_s(M_z)$ from differ. shapes  ~ 0.112  (smaller than analyses using hadr. Models)
- $\alpha_0$ ~ 0.5, appears to be "universal" within about 20%
- still a problem with calculation of $B_W$
- reasonable agreement with results from deep inelastic scattering (HERA)

# Gluon splittings g → b$\bar{\text{b}}$ and g → c$\bar{\text{c}}$

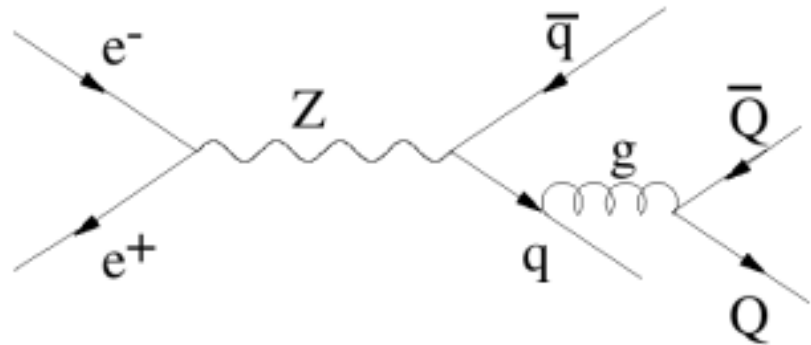

$$g_{b\bar{b}} = \frac{\text{BR}(Z \to q\bar{q}b\bar{b})}{\text{BR}(Z \to \text{hadrons})}$$

$$g_{4b} = \frac{\text{BR}(Z \to b\bar{b}b\bar{b})}{\text{BR}(Z \to \text{hadrons})}$$

$$g_{c\bar{c}} = \frac{\text{BR}(Z \to q\bar{q}c\bar{c})}{\text{BR}(Z \to \text{hadrons})}$$

|        | $g_{b\bar{b}} \times 10^3$ | $g_{4b} \times 10^4$ | $g_{c\bar{c}} \times 10^2$ |
|--------|---------------------------|----------------------|----------------------------|
| ALEPH  | $2.77 \pm 0.42 \pm 0.57$  |                      | $3.23 \pm 0.48 \pm 0.53$   |
| DELPHI | $3.3 \pm 1.0 \pm 0.8$     | $6.0 \pm 1.9 \pm 1.4$|                            |
| L3     |                           |                      | $2.45 \pm 0.29 \pm 0.53$   |
| OPAL   | $3.07 \pm 0.53 \pm 0.97$  | $3.6 \pm 1.7 \pm 2.7$| $3.20 \pm 0.21 \pm 0.38$   |
| SLD    | $2.84 \pm 0.61 \pm 0.59$  |                      |                            |
| Theory | 1.8-2.9                   | 3.2-5.2              | 1.3-2.0                    |

(H. Stenzel)

# Running b-quark mass: $m_b(M_Z)$

$R_b$: ratio of 3-jet rates of b- and light quark events

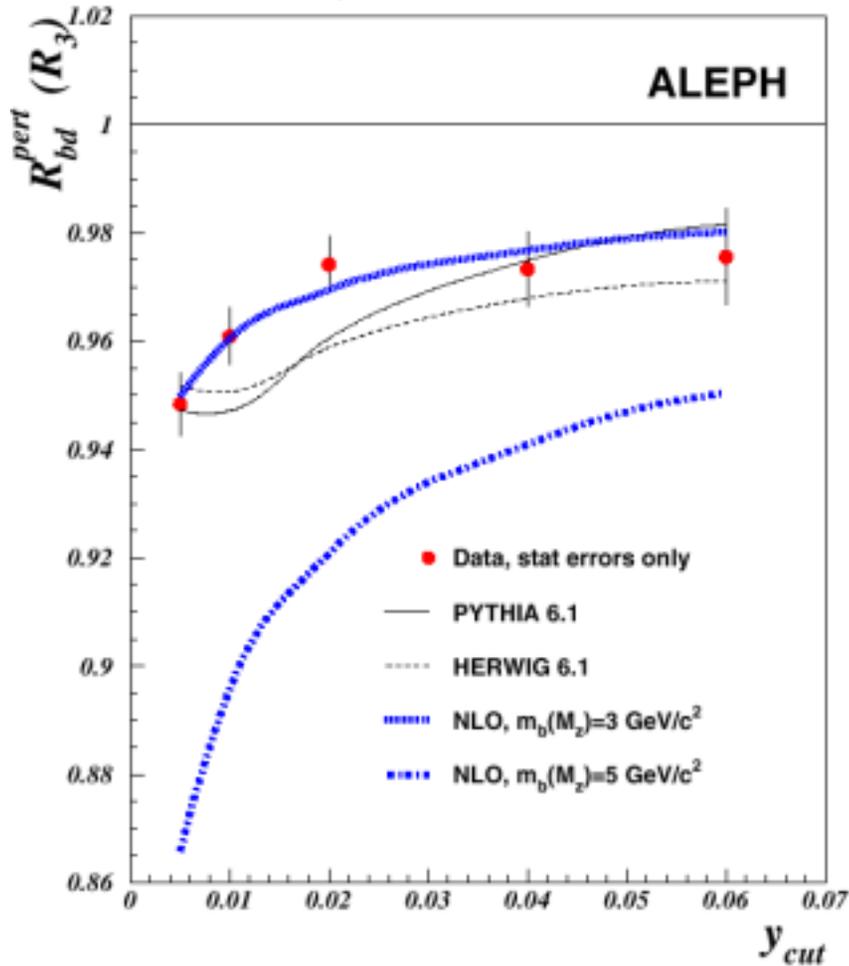
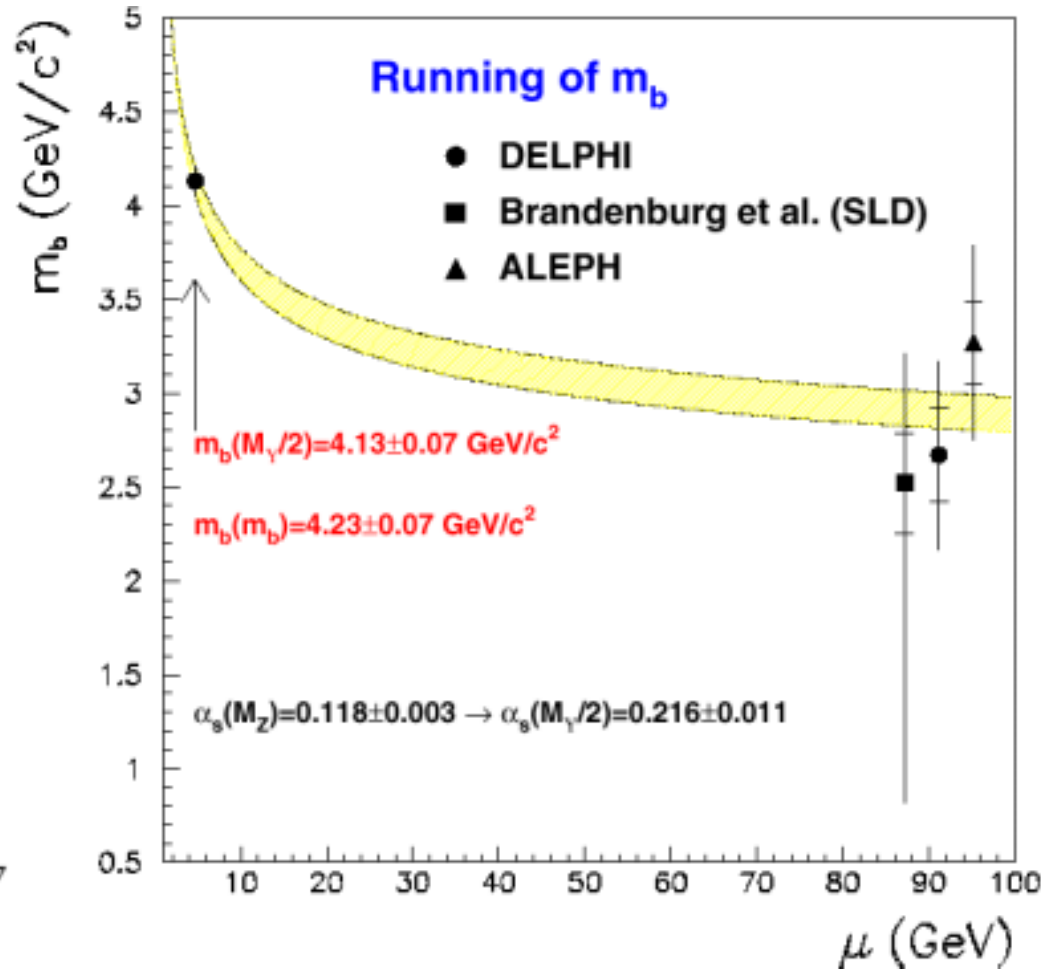

ALEPH: $m_b(M_Z) = 3.27 \pm 0.52$ GeV
DELPHI: $m_b(M_Z) = 2.61 \pm 0.54$ GeV

# Flavour independence of $\alpha_s$

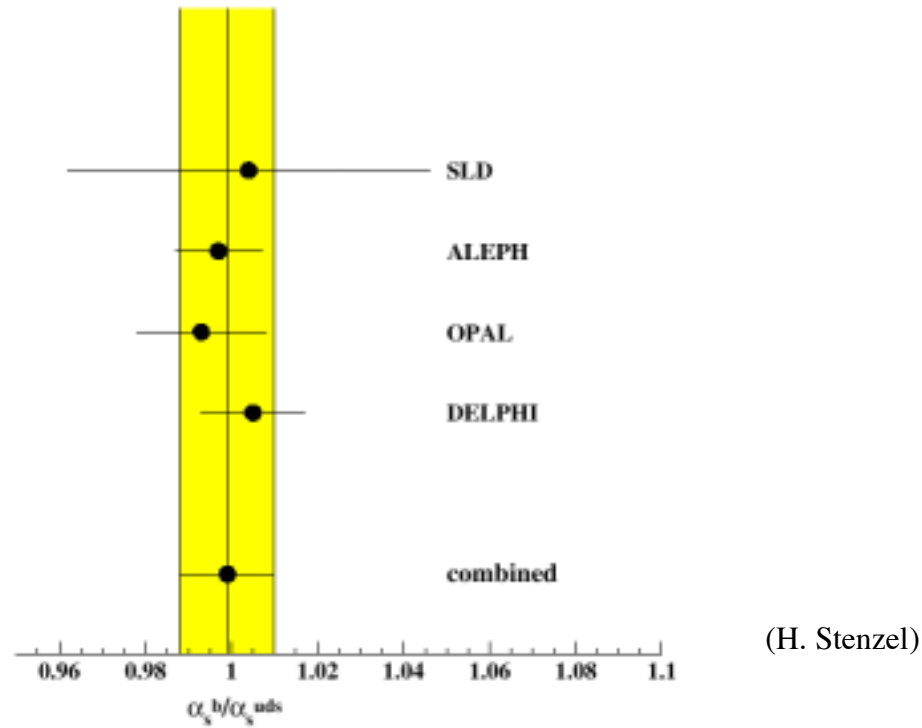

(H. Stenzel)

$\alpha_s^b / \alpha_s^{uds} = 0.999 \pm 0.011$ (combined SADO)

$\alpha_s^c / \alpha_s^{uds} = 1.012 \pm 0.040$ (combined SO)

$\alpha_s^s / \alpha_s^d = 0.956 \pm 0.053$ (OPAL $\langle n_{ch} \rangle$)

$\alpha_s^s / \alpha_s^u = 1.090 \pm 0.056$ (OPAL $\langle n_{ch} \rangle$)

$\alpha_s^u / \alpha_s^d = 0.877 \pm 0.081$ (OPAL $\langle n_{ch} \rangle$)

# 2-photon physics: $F_2^\gamma(x, Q^2)$

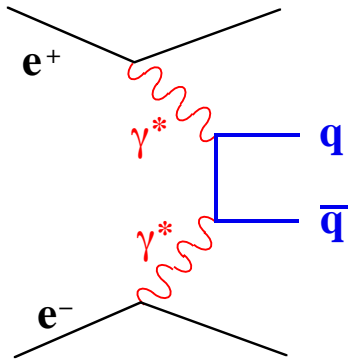

- $\dfrac{d^2\sigma_{e\gamma \to eX}}{dx\, dQ^2} = \dfrac{2\pi}{xQ^2}\left[\left(1 + (1-y)^2\right)F_2^\gamma(x, Q^2) - y^2 F_L^\gamma(x, Q^2)\right]$
- determined from single tag events
- $Q^2$ from scattered electron; $x$ from hadronic final state
- evidence for rise at low $x$ ?

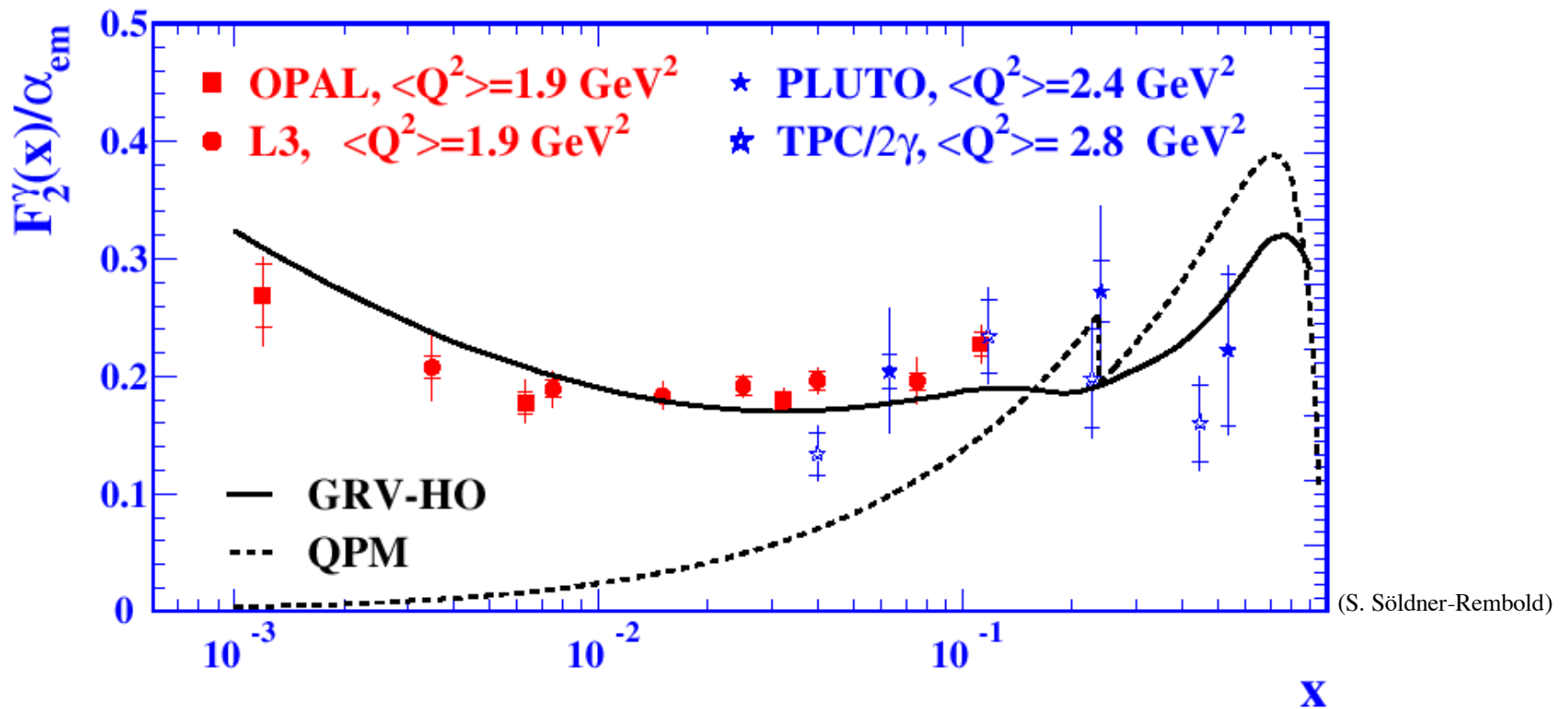

(S. Söldner-Rembold)

# 2-photon physics: scaling violations of $F_2^\gamma(x, Q^2)$

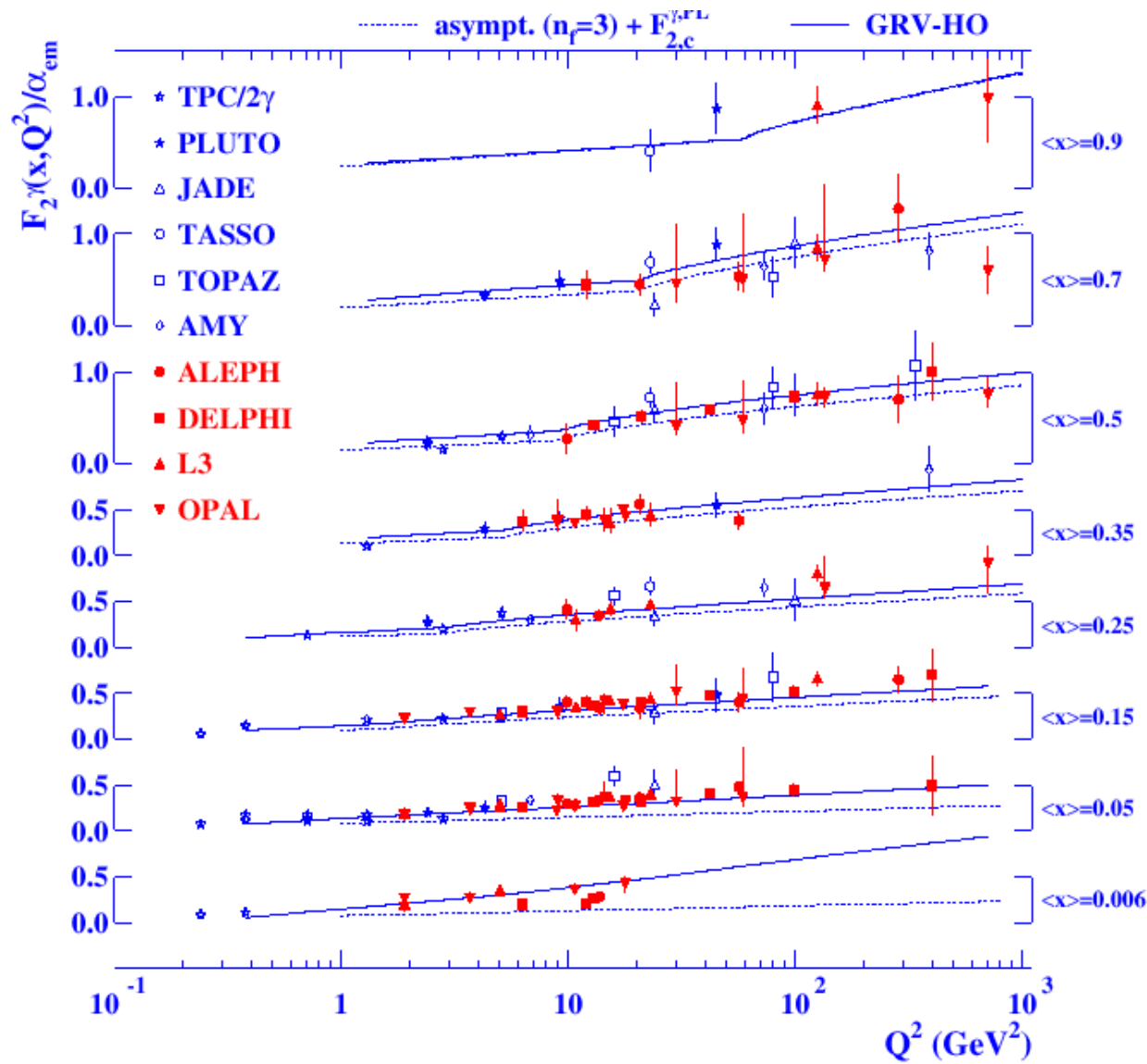

(S. Söldner-Rembold)

# 2-photon physics: heavy quark production

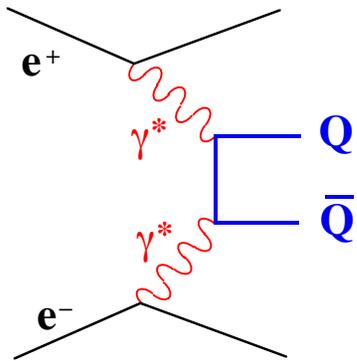
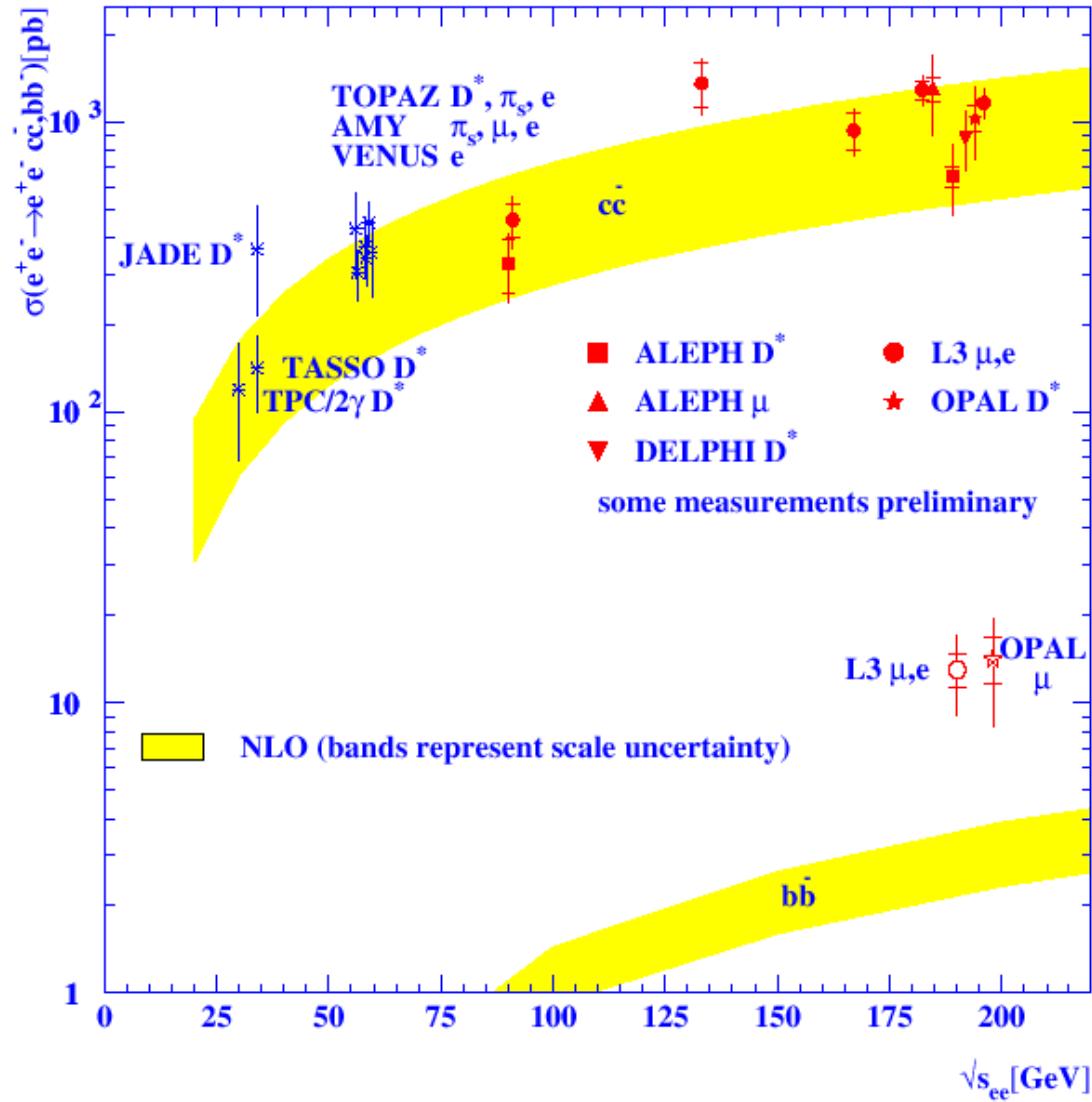

# Summary: QCD at LEP

- $\alpha_s(M_Z) = 0.121 \pm 0.005$ (resummed NLO; jets & shapes)
- $\alpha_s(M_Z) = 0.120 \pm 0.003$ (NNLO; $R_Z$, $R_\tau$) (World average $0.1184 \pm 0.0031$)
- running of $\alpha_s$, asymptotic freedom confirmed
- non-Abelian structure (gluon self-coupling) confirmed
- quark / gluon differences studied in detail
- effects of gluon coherence confirmed
- deeper understanding of hadronisation, in terms of power corrections, local parton-hadron duality, hadronisation models, ...

- running b-quark mass, flavour independence of $\alpha_s$ determined
- gluon splitting into heavy quarks
- 2-photon physics: $F_2^\gamma$ at small x, scaling violation, cross sections...

To do:
- NNLO for more observables (shapes; jets) !
  - deeper understanding of theoretical uncertainties
  - further assessment of non-perturbative effects